\documentclass[aps,prl,reprint,superscriptaddress,nofootinbib]{revtex4-2}
\usepackage{amsmath,amssymb,bm}
\usepackage{graphicx}
\usepackage[colorlinks=true,citecolor=blue,linkcolor=blue,urlcolor=blue]{hyperref}
\newcommand{\ii}{\mathrm{i}}
\begin{document}

\title{Emergent Conformal Symmetry on Superfluid Vortices\\in Two-Color QCD and Pseudoreal Gauge Theories}

\author{Muneto Nitta}
\affiliation{Department of Physics \& Research and Education Center for Natural Sciences, Keio University, 4-1-1 Hiyoshi, Yokohama, Kanagawa 223-8521, Japan}
\affiliation{International Institute for Sustainability with Knotted Chiral Meta Matter (WPI-SKCM$^2$), Hiroshima University, 1-3-2 Kagamiyama, Higashi-Hiroshima, Hiroshima 739-8511, Japan}
\date{\today}

\begin{abstract}
Two-color QCD provides a unique first-principles laboratory for strongly
interacting matter at finite baryon density, where a diquark condensate realizes
a baryonic superfluid.  We investigate the internal physics of its quantized
superfluid vortices in the chiral limit.  We find that a vortex localizes an exact
and normalizable $S^3\simeq SU(2)$ family of pion modes.  Remarkably, the bulk
Wess--Zumino--Witten (WZW) term induces a quantized WZW term for these modes,
driving the vortex theory in the infrared to the $SU(2)_1$ WZW conformal field
theory with central charge $c=1$.  Thus a strongly coupled conformal theory
emerges on a vortex in a nonconformal bulk theory.  The mechanism extends to
pseudoreal $Sp(2N)$ gauge theories, where the corresponding minimal vortex theory
flows to $SU(2)_N$, with the level directly encoding the microscopic anomaly
coefficient.
\end{abstract}

\maketitle

Two-color QCD (QC$_2$D) provides a rare first-principles laboratory for strongly
coupled matter at finite baryon density.  Pseudoreality removes the ordinary sign
problem, and at low temperature a diquark condensate breaks $U(1)_B$, producing a
baryonic superfluid \cite{Kogut1999,Kogut2000,SplittorffSonStephanov,
SplittorffNLO,Adhikari2018}.  Lattice calculations now resolve its phase structure,
equation of state, BEC--BCS crossover, and excitation spectrum
\cite{Hands2006,IidaItouLee2020,Boz2020,Begun2022,IidaItou2022,IidaItou2024,
ItouIida2024,ItouReview2025,ItouReview2026,IidaItou2026}.  Closely related
pseudoreal theories, including $SU(2)\simeq Sp(2)$ and $Sp(4)$ gauge theories, are
also widely studied as composite dark sectors
\cite{Hietanen2014,Hochberg2014,Hochberg2015,Bennett2018,Bennett2019,
BennettSinglet2024,KulkarniEtAl,ZierlerStrongDM}.

Quantized vortices are intrinsic to superfluid physics.  Rotation nucleates vortex
arrays, while vortex motion and interactions govern collective response and quantum
turbulence, as familiar from $^4$He, $^3$He, and ultracold atomic gases
\cite{Sonin1987,SalomaaVolovik1987,Fetter2009,TsubotaKasamatsu2025}.  Vortices are
likewise central to dense three-color QCD, in particular in connection with rotating
compact stars \cite{AlfordReview2008,EtoDenseReview2014,AlfordBaymFukushima2019}.
Yet the internal structure and quantum dynamics of superfluid vortices in QC$_2$D
remain largely unexplored.  Since QC$_2$D is accessible to finite-density lattice
simulations, its vortex sector offers an unusual opportunity to study defect physics
in strongly coupled baryonic matter from first principles.

Here we uncover a simple but unexpected structure in the chiral limit.  A baryonic
vortex localizes pion degrees of freedom into an exact, normalizable
$S^3\simeq SU(2)$ internal space.  More importantly, the microscopic QCD anomaly descends onto the vortex as a
quantized Wess--Zumino--Witten (WZW) term \cite{WessZumino1971,Witten1983}.
Related anomaly physics on defects appears in the Callan--Harvey inflow mechanism
\cite{CallanHarvey1985} and, in QCD chiral effective theory, in anomaly-induced
currents on axial vortices from the gauged WZW action \cite{FukushimaImaki2018}.
The present mechanism is distinct: the anomaly fixes a WZW term for dynamical,
normalizable internal vortex modes and thereby determines their infrared CFT.  The resulting relativistic
$1+1$-dimensional theory flows in the infrared to an
$SU(2)$ WZW conformal field theory (CFT) \cite{WittenBosonization1984,KZ1984,
BPZ1984,DiFrancesco1997}.  For a minimal QC$_2$D vortex the level is one and the central charge is
$c=1$; for pseudoreal $Sp(2N)$ gauge theories the level records the microscopic
anomaly coefficient.  Thus conformal dynamics emerges on a topological defect even
though the surrounding bulk theory is not conformal.

We first focus on two-color QCD and formulate it as the $N=1$ member of a
pseudoreal $Sp(2N)$ gauge theory with two massless Dirac flavors, with
$Sp(2)\simeq SU(2)$.  The chiral symmetry breaks $SU(4)\to Sp(4)$, and the
Nambu--Goldstone target is $SU(4)/Sp(4)\simeq S^5$.  It is convenient to write
a unit six-vector as
\begin{equation}
 N^A=\bigl(\pi^1,\pi^2,\pi^3,\sigma,\Re\Delta,\Im\Delta\bigr),\qquad N^A N^A=1 .
 \label{eq:Ndef}
\end{equation}
Here $A=1,\ldots,6$, $\bm\pi$ is the pion triplet, $\sigma$ the chiral direction, and $\Delta$
the complex diquark.  Schematically, their microscopic quark-bilinear realizations are
\begin{equation}
 \sigma\sim\bar\psi\psi,\quad \pi^a\sim\bar\psi\,\ii\gamma_5\tau^a\psi,\quad \Delta\sim\psi^T C\gamma_5\tau_2\epsilon_c\psi .
 \label{eq:bilinears}
\end{equation}
where $\psi$ is the Dirac quark field, $C$ is the charge-conjugation matrix,
$\tau^a$ are Pauli matrices in flavor space, and $\epsilon_c$ is the antisymmetric
invariant tensor of two colors.  The leading-order kinetic Lagrangian reads
\begin{align}
 {\cal L}_2&={f_\pi^2\over2}
 \left[(\partial_\mu\sigma)^2+(\partial_\mu\bm\pi)^2
 +|D_\mu\Delta|^2\right],\nonumber\\
 D_0\Delta&=(\partial_0-\ii\mu_B)\Delta,\qquad
 D_i\Delta=\partial_i\Delta .
 \label{eq:LO}
\end{align}
Here $\mu_B$ is the baryon chemical potential; we
normalize the baryon charge of the diquark field $\Delta$ to unity
\cite{Kogut1999,Kogut2000}.  The chemical potential selects the baryonic two-plane and leaves
$SO(4)\simeq SU(2)_L\times SU(2)_R$ acting on $(\sigma,\bm\pi)$.
The homogeneous minimum has $|\Delta|=1$ and $\sigma=\bm\pi=0$.
For general even $N_f>2$, an antisymmetric diquark condensate further breaks
$SU(N_f)_L\times SU(N_f)_R$ to $Sp(N_f)_L\times Sp(N_f)_R$
\cite{Kogut2000,KanazawaWettigYamamoto2009}, while at $N_f=2$ this extra
breaking is absent since $Sp(2)\simeq SU(2)$.

The effective action also contains the bulk WZW term for $Sp(2N)$ fundamentals,
\begin{equation}
 S_{\rm WZW}=2\pi\ii\,N\int_{M_5}\Omega_5 ,
 \label{eq:WZW}
\end{equation}
where our convention is that the fundamental representation of $Sp(2N)$ has
dimension $2N$, and $\Omega_5$ is the normalized generator of $H^5(S^5,\mathbb Z)$,
\begin{align}
 \Omega_5={}&{1\over 5!\,\pi^3}\,\epsilon_{ABCDEF}N^A
 dN^B\wedge dN^C\wedge dN^D\nonumber\\
 &{}\wedge dN^E\wedge dN^F,\qquad
 \int_{S^5}\Omega_5=1 .
 \label{eq:Omega5}
\end{align}
Its normalization is fixed by the microscopic flavor anomaly
\cite{BraunerKolesova2019,LeeOhmoriTachikawa,Saito}.  The chemical potential may
equivalently be viewed as a flat $U(1)_B$ background gauge field.  For the straight-vortex
ansatz with $n=n(t,z)$ and a flat $U(1)_B$ background, the gauged WZW completion
gives no additional contribution to the $1+1$-dimensional orientational action at this
derivative order \cite{BraunerKolesova2019}.  For the static
vortex ansatz below with a fixed internal orientation, the WZW term does not modify
the classical profile equation.  It becomes nontrivial when the internal orientation
is promoted to a field on the vortex worldsheet.

\begin{figure}[t]
 \includegraphics[width=.88\columnwidth]{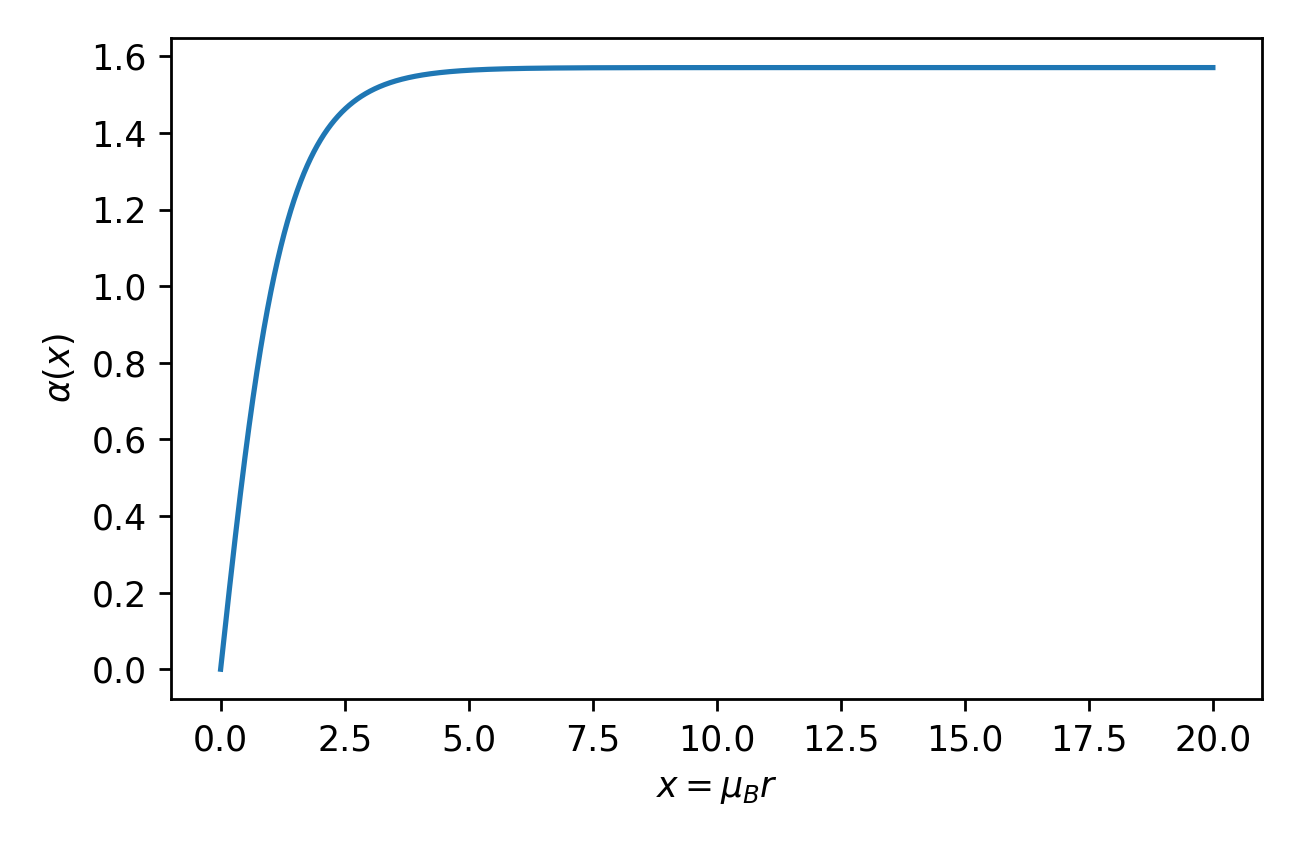}
 \caption{Chiral-limit unit-vortex profile as a function of $x=\mu_B r$.
 The $S^3$ radius in field space is $\cos\alpha(x)$ and therefore collapses
 exponentially in the homogeneous bulk.}
 \label{fig:profile}
\end{figure}

For a straight vortex of winding number $\nu\in\mathbb Z$ along the $z$ axis, we take the ansatz
\begin{align}
 \Delta&=\sin\alpha(r)e^{\ii\nu\theta},\qquad
 (\sigma,\bm\pi)=\cos\alpha(r)\,n,\nonumber\\
 n&=(n^0,n^1,n^2,n^3)\in\mathbb R^4,\qquad n^\alpha n^\alpha=1 .
 \label{eq:vortex}
\end{align}
Here $\alpha=0,1,2,3$ and $n^\alpha$ is a real four-component unit vector specifying the orientation of $(\sigma,\bm\pi)$; hence it parametrizes $S^3$.  The boundary conditions are $\alpha(0)=0$ and $\alpha(\infty)=\pi/2$.
Thus, for constant $n$, the radial equation is
\begin{equation}
 \alpha''+{\alpha'\over r}
 -{\nu^2\over r^2}\sin\alpha\cos\alpha
 +\mu_B^2\sin\alpha\cos\alpha=0 ,
 \label{eq:alpha}
\end{equation}
where a prime denotes $d/dr$.  A numerical solution for a unit vortex ($\nu=1$) is shown in Fig.~\ref{fig:profile}.

All values of $n$ have exactly the same energy.
The bulk vacuum is invariant under $SO(4)$ because the coefficient of $n$ vanishes
as $r\to\infty$, while a particular vortex core chooses one $n$.  Thus a particular vortex spontaneously breaks the unbroken bulk symmetry according to
\begin{align}
 SO(4)&\longrightarrow SO(3),\nonumber\\
 {\cal M}_{\rm vortex}&\equiv {SO(4)\over SO(3)}\simeq S^3\simeq SU(2).
 \label{eq:moduli}
\end{align}
where the arrow denotes the symmetry breaking induced by choosing a core orientation, and
${\cal M}_{\rm vortex}$ is the corresponding internal moduli space of a single vortex.
This is a genuine symmetry-generated moduli space, not merely the $S^3$ slice
allowed by the constraint at $r=0$.  The distinction matters physically.  The
three pion directions are ordinary bulk fields, but the vortex multiplies them by
a radial envelope $\cos\alpha(r)$ that vanishes in the homogeneous superfluid.
Thus the internal orientation can fluctuate along the string without rotating the
bulk condensate.  In this sense the vortex converts bulk pion directions into
localized collective coordinates, a possibility made especially clean at $N_f=2$
by the absence of additional bulk flavor breaking.

Normalizability follows from the same geometry.  With $x=\mu_B r$ and
$\beta=\pi/2-\alpha$, the asymptotic equation is
\begin{equation}
 \beta_{xx}+{1\over x}\beta_x+
 \left({\nu^2\over x^2}-1\right)\beta=0 ,
\end{equation}
so $\beta\propto K_{\ii\nu}(x)\sim e^{-x}/\sqrt{x}$.  Hence
$\cos\alpha$ decays exponentially and the internal zero modes have finite norm.
In the moduli-space approximation, we promote the internal orientation to a slowly varying worldsheet field, $n=n(t,z)$ \cite{Manton1982,EtoEffective2006}.  Writing the worldsheet coordinates as $x^\gamma=(t,z)$, substitution into the bulk kinetic term gives
\begin{align}
 S_{\rm kin}^{\rm ws}&={Z_\nu\over2}\int d^2x\,(\partial_\gamma n^\alpha)(\partial^\gamma n^\alpha),\nonumber\\
 Z_\nu&=2\pi f_\pi^2\int_0^\infty r\,dr\,\cos^2\alpha(r)={\pi\nu^2f_\pi^2\over\mu_B^2}.
 \label{eq:exactmetric}
\end{align}
Remarkably, the second equality fixes the coefficient without solving the profile and follows from the Pohozaev identity derived in the End Matter.
The induced metric is therefore the round bi-invariant metric on $S^3$ with an
exactly determined overall coefficient.  This is stronger than a generic
collective-coordinate result: no numerical overlap integral is needed once the
vortex solves the LO equation.  The scale of the internal theory is set by
$f_\pi^2/\mu_B^2$, while its geometry is protected by the unbroken $SO(4)$.
The numerical profile in Fig.~\ref{fig:profile} also verifies Eq.~(\ref{eq:exactmetric}).
We now evaluate the bulk WZW term in Eq.~(\ref{eq:WZW}) on the vortex.  In the coordinates
(\ref{eq:vortex}),
\begin{equation}
 \Omega_5={\nu\over\pi^3}\,
 \alpha'\sin\alpha\cos^3\alpha\,
 dr\wedge d\theta\wedge{\rm vol}_{S^3}.
\end{equation}
Using only $\alpha(0)=0$ and $\alpha(\infty)=\pi/2$,
\begin{equation}
 \int_{D_\perp^2}\Omega_5
 =\nu\,\Omega_3,\qquad
 \Omega_3={{\rm vol}_{S^3}\over2\pi^2},\quad
 \int_{S^3}\Omega_3=1 .
 \label{eq:descent}
\end{equation}
Thus
\begin{equation}
 S_{\rm WZW}^{\rm ws}=2\pi\ii\,N\nu\int_{B_3}\Omega_3,\qquad
 k_{\rm ws}=N\nu .
 \label{eq:level}
\end{equation}
Unlike the metric coefficient, whose value follows from the vortex dynamics,
the integer level is fixed by the microscopic flavor anomaly and the vortex
winding.  Equation~(\ref{eq:descent}) also gives a direct geometric picture of
the anomaly descent.  The five-dimensional extension factorizes locally as
$D_\perp^2\times B_3$: the transverse disk carries the superfluid winding and
$B_3$ is the WZW extension of the $S^3$ worldsheet field.  The bulk $\pi_5$
winding is therefore resolved into the vortex winding times the worldsheet
$S^3$ winding.\footnote{A lower-dimensional analogue occurs in finite-isospin QCD: in the chiral limit, a charged-pion vortex can localize a neutral-pion phase, whose winding along the vortex realizes the bulk $\pi_3$ charge \cite{QiuNitta2024}.  At finite pion mass, related isospin-QCD phases realize baryon number through linked vortices \cite{QiuNitta2025,QiuNitta2026}.}

Combining Eqs.~(\ref{eq:exactmetric}) and (\ref{eq:level}), the chiral-limit
worldsheet theory is an $SU(2)$ principal chiral model with a level-$N\nu$ WZW term.  Identifying the unit four-vector with the group element
$g=n^0\mathbf 1+\ii n^a\tau^a\in SU(2)$ makes the equivalence $S^3\simeq SU(2)$ explicit:
\begin{equation}
 S_{\rm ws}={Z_\nu\over2}\int d^2x\,(\partial_\gamma n^\alpha)(\partial^\gamma n^\alpha)
 +2\pi\ii N\nu\int_{B_3}\Omega_3 .
 \label{eq:ws}
\end{equation}
At the two-derivative order considered here, the $S^3$ orientation is orthogonal
in the $S^5$ target-space metric to the $U(1)_B$ phase, so the localized
internal sector does not mix kinetically with the gapless bulk superfluid phonon.
The bare sigma-model coupling need not coincide with its conformal value at the
matching scale.  Under the two-dimensional RG flow, the principal chiral model
with WZW level $k$ flows to the $SU(2)_k$ WZW fixed point
\cite{PolyakovWiegmann1984,SchubringShifman2020}.  Coleman's theorem on continuous symmetry breaking in $1+1$ dimensions
\cite{Coleman1973} is fully respected: the classical $SO(4)\to SO(3)$
long-range order is restored quantum mechanically.  What is exceptional is that
symmetry restoration does \emph{not} generate a mass gap.  Instead the quantized
WZW term protects a gapless infrared endpoint whose current algebra, central charge,
and scaling dimensions are those of the exact $SU(2)_k$ CFT.
Consequently, with $k=N|\nu|$, the Sugawara construction gives
$c=k\,\dim SU(2)/(k+h^\vee)=3k/(k+2)$, where $h^\vee=2$ is the dual Coxeter number of $SU(2)$ \cite{KZ1984,DiFrancesco1997}.  Thus
\begin{equation}
 {\rm IR}:\qquad SU(2)_{N|\nu|},\qquad
 c={3N|\nu|\over N|\nu|+2}.
 \label{eq:CFT}
\end{equation}
For a minimal vortex, $|\nu|=1$, this gives $c=3N/(N+2)$.  In particular, for QC$_2$D, $N=1$, the infrared theory is $SU(2)_1$ with central charge $c=1$.  For $Sp(4)$ the corresponding minimal vortex has level two and flows to
$SU(2)_2$.  The common local target geometry therefore retains microscopic
gauge-group information through the anomaly coefficient.  This provides a rare
example in which the local moduli-space geometry is identical across a family of
theories while the vortex CFT distinguishes the microscopic gauge representation.

It is useful at this point to compare this infrared behavior with other vortex
worldsheet theories.  Non-Abelian vortices in supersymmetric gauge
theories possess $\mathbb{C}P^{N-1}$-type orientational modes
\cite{HananyTong2003,AuzziEtAl2003,EtoEtAl2006,EtoHigherWinding2006,
TongTASI2005,EtoReview2006,ShifmanYungRMP2007,ShifmanYungBook2009,TongReview2009}.
Their quantum worldsheet dynamics is asymptotically free and generates a mass
scale \cite{ShifmanYungMonopoles2004,HananyTongQuantum2004}, consistently with
Coleman's theorem.  A closely related story occurs for non-Abelian superfluid
vortices in the color-flavor-locked (CFL) phase of dense QCD: the vortex solutions
\cite{BalachandranDigalMatsuura2006,EtoNitta2009}, their normalizable $\mathbb{C}P^2$
orientational zero modes \cite{NakanoNittaMatsuura2008}, and the corresponding
worldsheet theory \cite{EtoNakanoNitta2009,EtoNittaYamamoto2010} have been established explicitly,
while quantum $\mathbb{C}P^2$ dynamics generates a nonperturbative mass scale
\cite{GorskyShifmanYung2011,EtoNittaYamamoto2011}.
Other routes to gapless defect dynamics are known: nonrelativistic vortices can
evade the relativistic infrared behavior through quadratically dispersing
Nambu--Goldstone modes \cite{NittaUchinoVinci2014}, while supersymmetric vortices
at specially tuned Argyres--Douglas points can flow to superconformal theories
\cite{TongSC}.  The present mechanism is different from both.  It requires
neither nonrelativistic dynamics nor supersymmetry or a conformal bulk: the
quantized WZW term inherited from the QCD anomaly fixes the gapless infrared
endpoint.  The $SU(2)_{N|\nu|}$ fixed point therefore governs the long-distance
dynamics of the localized internal sector of an isolated vortex in the chiral
limit.  Higher-derivative density effects may renormalize nonuniversal quantities
such as the worldsheet velocity, while finite quark mass, temperature, finite
size, and intervortex interactions provide physical perturbations away from this
critical regime.

A finite quark mass qualitatively changes the present construction.  The
homogeneous QC$_2$D condensate then contains a nonzero chiral component,
$\cos\alpha_\infty=m_\pi^2/\mu_B^2$, so a uniform $S^3$ rotation changes the
field all the way to infinity and is not normalizable.  The exact $S^3$ moduli
and the derivation of Eq.~(\ref{eq:level}) as a physical worldsheet WZW term
therefore belong specifically to the chiral limit.  Nevertheless, the vortex
background can still support normalizable massive pion bound states.  The
finite-mass problem should thus be regarded as a distinct deformation away from
the exact moduli problem, whose nonlinear effective dynamics and anomaly-induced
interactions remain to be determined.

The chiral limit thus reveals a direct route by which four-dimensional anomaly
data determine the infrared universality class of a topological defect.  For a
minimal vortex, QC$_2$D realizes the $SU(2)_1$ WZW CFT, while pseudoreal
$Sp(2N)$ gauge theories generate the anomaly-determined sequence $SU(2)_N$.
Intriguingly, the QC$_2$D vortex is in the same infrared universality class as the
spin-$1/2$ antiferromagnetic Heisenberg chain
\cite{AffleckHaldane1987,ItoiMukaida1994}; established results for its finite-size
spectrum and correlation functions therefore give concrete universal predictions
for long-distance observables on the vortex \cite{AffleckGepnerSchulzZiman1989}.
The parallel goes beyond a coincidence of universality classes.  Half-odd-integer
spin chains obey Lieb--Schultz--Mattis-type constraints that obstruct a unique,
symmetric, trivially gapped ground state \cite{LSM1961,Oshikawa2000,Tasaki2022},
whereas here nontrivial infrared dynamics is enforced by the four-dimensional
anomaly descending to the vortex WZW term.  The microscopic origins of the two
constraints are different, but both lead naturally to robust nontrivial infrared
physics.

An important next step is a first-principles test of these predictions.  Vortices
have already been generated in nonperturbative lattice-field simulations of
rotating superfluids and gauge theories \cite{HayataYamamoto2015,Yamamoto2018,YamamotoHirono2013}.
Introducing a quantized vortex in finite-density QC$_2$D and measuring pion
correlation functions along its core would directly probe the predicted
$SU(2)_1$ critical behavior; finite-size scaling could determine its scaling
dimensions and central charge $c=1$.  Another intriguing direction is a sufficiently
large or stabilized vortex ring, for which the worldsheet CFT is placed on a spatial
circle.  Its finite-size spectrum should organize into the conformal towers of the
$SU(2)_1$ WZW theory, providing direct access to scaling dimensions and the central
charge through universal finite-size effects
\cite{AffleckGepnerSchulzZiman1989,BloeteCardyNightingale1986}.  More generally,
simulations of pseudoreal $Sp(2N)$ gauge theories could test the anomaly-determined
sequence $SU(2)_N$, with $c=3N/(N+2)$ for a minimal vortex.  This offers a rare
opportunity to realize and test an exactly characterized two-dimensional CFT as a
localized defect sector of a four-dimensional strongly coupled gauge theory,
providing a concrete bridge from QCD-like gauge dynamics to exactly solvable CFTs
through topological defects.

\begin{acknowledgments}
This work is supported in part by Japan Society for the Promotion of Science (JSPS) KAKENHI [Grants No.~JP22H01221 and JP23K22492] and the WPI program ``Sustainability with Knotted Chiral Meta Matter (WPI-SKCM$^2$)'' at Hiroshima University.
\end{acknowledgments}

\bibliographystyle{apsrev4-2}
\bibliography{qc2d_vortex_references_v44}

\begin{thebibliography}{80}%
\makeatletter
\providecommand \@ifxundefined [1]{%
 \@ifx{#1\undefined}
}%
\providecommand \@ifnum [1]{%
 \ifnum #1\expandafter \@firstoftwo
 \else \expandafter \@secondoftwo
 \fi
}%
\providecommand \@ifx [1]{%
 \ifx #1\expandafter \@firstoftwo
 \else \expandafter \@secondoftwo
 \fi
}%
\providecommand \natexlab [1]{#1}%
\providecommand \enquote  [1]{``#1''}%
\providecommand \bibnamefont  [1]{#1}%
\providecommand \bibfnamefont [1]{#1}%
\providecommand \citenamefont [1]{#1}%
\providecommand \href@noop [0]{\@secondoftwo}%
\providecommand \href [0]{\begingroup \@sanitize@url \@href}%
\providecommand \@href[1]{\@@startlink{#1}\@@href}%
\providecommand \@@href[1]{\endgroup#1\@@endlink}%
\providecommand \@sanitize@url [0]{\catcode `\\12\catcode `\$12\catcode
  `\&12\catcode `\#12\catcode `\^12\catcode `\_12\catcode `\%12\relax}%
\providecommand \@@startlink[1]{}%
\providecommand \@@endlink[0]{}%
\providecommand \url  [0]{\begingroup\@sanitize@url \@url }%
\providecommand \@url [1]{\endgroup\@href {#1}{\urlprefix }}%
\providecommand \urlprefix  [0]{URL }%
\providecommand \Eprint [0]{\href }%
\providecommand \doibase [0]{https://doi.org/}%
\providecommand \selectlanguage [0]{\@gobble}%
\providecommand \bibinfo  [0]{\@secondoftwo}%
\providecommand \bibfield  [0]{\@secondoftwo}%
\providecommand \translation [1]{[#1]}%
\providecommand \BibitemOpen [0]{}%
\providecommand \bibitemStop [0]{}%
\providecommand \bibitemNoStop [0]{.\EOS\space}%
\providecommand \EOS [0]{\spacefactor3000\relax}%
\providecommand \BibitemShut  [1]{\csname bibitem#1\endcsname}%
\let\auto@bib@innerbib\@empty
\bibitem [{\citenamefont {Kogut}\ \emph {et~al.}(1999)\citenamefont {Kogut},
  \citenamefont {Stephanov},\ and\ \citenamefont {Toublan}}]{Kogut1999}%
  \BibitemOpen
  \bibfield  {author} {\bibinfo {author} {\bibfnamefont {J.~B.}\ \bibnamefont
  {Kogut}}, \bibinfo {author} {\bibfnamefont {M.~A.}\ \bibnamefont
  {Stephanov}},\ and\ \bibinfo {author} {\bibfnamefont {D.}~\bibnamefont
  {Toublan}},\ }\href {https://doi.org/10.1016/S0370-2693(99)00971-5}
  {\bibfield  {journal} {\bibinfo  {journal} {Phys. Lett. B}\ }\textbf
  {\bibinfo {volume} {464}},\ \bibinfo {pages} {183} (\bibinfo {year}
  {1999})},\ \Eprint {https://arxiv.org/abs/hep-ph/9906346}
  {arXiv:hep-ph/9906346} \BibitemShut {NoStop}%
\bibitem [{\citenamefont {Kogut}\ \emph {et~al.}(2000)\citenamefont {Kogut},
  \citenamefont {Stephanov}, \citenamefont {Toublan}, \citenamefont
  {Verbaarschot},\ and\ \citenamefont {Zhitnitsky}}]{Kogut2000}%
  \BibitemOpen
  \bibfield  {author} {\bibinfo {author} {\bibfnamefont {J.~B.}\ \bibnamefont
  {Kogut}}, \bibinfo {author} {\bibfnamefont {M.~A.}\ \bibnamefont
  {Stephanov}}, \bibinfo {author} {\bibfnamefont {D.}~\bibnamefont {Toublan}},
  \bibinfo {author} {\bibfnamefont {J.~J.~M.}\ \bibnamefont {Verbaarschot}},\
  and\ \bibinfo {author} {\bibfnamefont {A.}~\bibnamefont {Zhitnitsky}},\
  }\href {https://doi.org/10.1016/S0550-3213(00)00242-X} {\bibfield  {journal}
  {\bibinfo  {journal} {Nucl. Phys. B}\ }\textbf {\bibinfo {volume} {582}},\
  \bibinfo {pages} {477} (\bibinfo {year} {2000})},\ \Eprint
  {https://arxiv.org/abs/hep-ph/0001171} {arXiv:hep-ph/0001171} \BibitemShut
  {NoStop}%
\bibitem [{\citenamefont {Splittorff}\ \emph {et~al.}(2001)\citenamefont
  {Splittorff}, \citenamefont {Son},\ and\ \citenamefont
  {Stephanov}}]{SplittorffSonStephanov}%
  \BibitemOpen
  \bibfield  {author} {\bibinfo {author} {\bibfnamefont {K.}~\bibnamefont
  {Splittorff}}, \bibinfo {author} {\bibfnamefont {D.~T.}\ \bibnamefont
  {Son}},\ and\ \bibinfo {author} {\bibfnamefont {M.~A.}\ \bibnamefont
  {Stephanov}},\ }\href {https://doi.org/10.1103/PhysRevD.64.016003} {\bibfield
   {journal} {\bibinfo  {journal} {Phys. Rev. D}\ }\textbf {\bibinfo {volume}
  {64}},\ \bibinfo {pages} {016003} (\bibinfo {year} {2001})},\ \Eprint
  {https://arxiv.org/abs/hep-ph/0012274} {arXiv:hep-ph/0012274} \BibitemShut
  {NoStop}%
\bibitem [{\citenamefont {Splittorff}\ \emph {et~al.}(2002)\citenamefont
  {Splittorff}, \citenamefont {Toublan},\ and\ \citenamefont
  {Verbaarschot}}]{SplittorffNLO}%
  \BibitemOpen
  \bibfield  {author} {\bibinfo {author} {\bibfnamefont {K.}~\bibnamefont
  {Splittorff}}, \bibinfo {author} {\bibfnamefont {D.}~\bibnamefont
  {Toublan}},\ and\ \bibinfo {author} {\bibfnamefont {J.~J.~M.}\ \bibnamefont
  {Verbaarschot}},\ }\href {https://doi.org/10.1016/S0550-3213(01)00536-3}
  {\bibfield  {journal} {\bibinfo  {journal} {Nucl. Phys. B}\ }\textbf
  {\bibinfo {volume} {620}},\ \bibinfo {pages} {290} (\bibinfo {year}
  {2002})},\ \Eprint {https://arxiv.org/abs/hep-ph/0108040}
  {arXiv:hep-ph/0108040} \BibitemShut {NoStop}%
\bibitem [{\citenamefont {Adhikari}\ \emph {et~al.}(2018)\citenamefont
  {Adhikari}, \citenamefont {Beleznay},\ and\ \citenamefont
  {Mannarelli}}]{Adhikari2018}%
  \BibitemOpen
  \bibfield  {author} {\bibinfo {author} {\bibfnamefont {P.}~\bibnamefont
  {Adhikari}}, \bibinfo {author} {\bibfnamefont {S.~B.}\ \bibnamefont
  {Beleznay}},\ and\ \bibinfo {author} {\bibfnamefont {M.}~\bibnamefont
  {Mannarelli}},\ }\href {https://doi.org/10.1140/epjc/s10052-018-5934-6}
  {\bibfield  {journal} {\bibinfo  {journal} {Eur. Phys. J. C}\ }\textbf
  {\bibinfo {volume} {78}},\ \bibinfo {pages} {441} (\bibinfo {year} {2018})},\
  \Eprint {https://arxiv.org/abs/1803.00490} {arXiv:1803.00490 [hep-th]}
  \BibitemShut {NoStop}%
\bibitem [{\citenamefont {Hands}\ \emph {et~al.}(2006)\citenamefont {Hands},
  \citenamefont {Kim},\ and\ \citenamefont {Skullerud}}]{Hands2006}%
  \BibitemOpen
  \bibfield  {author} {\bibinfo {author} {\bibfnamefont {S.}~\bibnamefont
  {Hands}}, \bibinfo {author} {\bibfnamefont {S.}~\bibnamefont {Kim}},\ and\
  \bibinfo {author} {\bibfnamefont {J.-I.}\ \bibnamefont {Skullerud}},\ }\href
  {https://doi.org/10.1140/epjc/s2006-02621-8} {\bibfield  {journal} {\bibinfo
  {journal} {Eur. Phys. J. C}\ }\textbf {\bibinfo {volume} {48}},\ \bibinfo
  {pages} {193} (\bibinfo {year} {2006})},\ \Eprint
  {https://arxiv.org/abs/hep-lat/0604004} {arXiv:hep-lat/0604004} \BibitemShut
  {NoStop}%
\bibitem [{\citenamefont {Iida}\ \emph {et~al.}(2020)\citenamefont {Iida},
  \citenamefont {Itou},\ and\ \citenamefont {Lee}}]{IidaItouLee2020}%
  \BibitemOpen
  \bibfield  {author} {\bibinfo {author} {\bibfnamefont {K.}~\bibnamefont
  {Iida}}, \bibinfo {author} {\bibfnamefont {E.}~\bibnamefont {Itou}},\ and\
  \bibinfo {author} {\bibfnamefont {T.-G.}\ \bibnamefont {Lee}},\ }\href
  {https://doi.org/10.1007/JHEP01(2020)181} {\bibfield  {journal} {\bibinfo
  {journal} {JHEP}\ }\textbf {\bibinfo {volume} {01}},\ \bibinfo {pages}
  {181}},\ \Eprint {https://arxiv.org/abs/1910.07872} {arXiv:1910.07872
  [hep-lat]} \BibitemShut {NoStop}%
\bibitem [{\citenamefont {Boz}\ \emph {et~al.}(2020)\citenamefont {Boz},
  \citenamefont {Giudice}, \citenamefont {Hands},\ and\ \citenamefont
  {Skullerud}}]{Boz2020}%
  \BibitemOpen
  \bibfield  {author} {\bibinfo {author} {\bibfnamefont {T.}~\bibnamefont
  {Boz}}, \bibinfo {author} {\bibfnamefont {P.}~\bibnamefont {Giudice}},
  \bibinfo {author} {\bibfnamefont {S.}~\bibnamefont {Hands}},\ and\ \bibinfo
  {author} {\bibfnamefont {J.-I.}\ \bibnamefont {Skullerud}},\ }\href
  {https://doi.org/10.1103/PhysRevD.101.074506} {\bibfield  {journal} {\bibinfo
   {journal} {Phys. Rev. D}\ }\textbf {\bibinfo {volume} {101}},\ \bibinfo
  {pages} {074506} (\bibinfo {year} {2020})},\ \Eprint
  {https://arxiv.org/abs/1912.10975} {arXiv:1912.10975 [hep-lat]} \BibitemShut
  {NoStop}%
\bibitem [{\citenamefont {Begun}\ \emph {et~al.}(2022)\citenamefont {Begun},
  \citenamefont {Bornyakov}, \citenamefont {Goy}, \citenamefont {Nakamura},\
  and\ \citenamefont {Rogalyov}}]{Begun2022}%
  \BibitemOpen
  \bibfield  {author} {\bibinfo {author} {\bibfnamefont {A.}~\bibnamefont
  {Begun}}, \bibinfo {author} {\bibfnamefont {V.~G.}\ \bibnamefont
  {Bornyakov}}, \bibinfo {author} {\bibfnamefont {V.~A.}\ \bibnamefont {Goy}},
  \bibinfo {author} {\bibfnamefont {A.}~\bibnamefont {Nakamura}},\ and\
  \bibinfo {author} {\bibfnamefont {R.~N.}\ \bibnamefont {Rogalyov}},\ }\href
  {https://doi.org/10.1103/PhysRevD.105.114505} {\bibfield  {journal} {\bibinfo
   {journal} {Phys. Rev. D}\ }\textbf {\bibinfo {volume} {105}},\ \bibinfo
  {pages} {114505} (\bibinfo {year} {2022})},\ \Eprint
  {https://arxiv.org/abs/2203.04909} {arXiv:2203.04909 [hep-lat]} \BibitemShut
  {NoStop}%
\bibitem [{\citenamefont {Iida}\ and\ \citenamefont
  {Itou}(2022)}]{IidaItou2022}%
  \BibitemOpen
  \bibfield  {author} {\bibinfo {author} {\bibfnamefont {K.}~\bibnamefont
  {Iida}}\ and\ \bibinfo {author} {\bibfnamefont {E.}~\bibnamefont {Itou}},\
  }\href {https://doi.org/10.1093/ptep/ptac137} {\bibfield  {journal} {\bibinfo
   {journal} {PTEP}\ }\textbf {\bibinfo {volume} {2022}},\ \bibinfo {pages}
  {111B01} (\bibinfo {year} {2022})},\ \Eprint
  {https://arxiv.org/abs/2207.01253} {arXiv:2207.01253 [hep-ph]} \BibitemShut
  {NoStop}%
\bibitem [{\citenamefont {Iida}\ \emph {et~al.}(2024)\citenamefont {Iida},
  \citenamefont {Itou}, \citenamefont {Murakami},\ and\ \citenamefont
  {Suenaga}}]{IidaItou2024}%
  \BibitemOpen
  \bibfield  {author} {\bibinfo {author} {\bibfnamefont {K.}~\bibnamefont
  {Iida}}, \bibinfo {author} {\bibfnamefont {E.}~\bibnamefont {Itou}}, \bibinfo
  {author} {\bibfnamefont {K.}~\bibnamefont {Murakami}},\ and\ \bibinfo
  {author} {\bibfnamefont {D.}~\bibnamefont {Suenaga}},\ }\href
  {https://doi.org/10.1007/JHEP10(2024)022} {\bibfield  {journal} {\bibinfo
  {journal} {JHEP}\ }\textbf {\bibinfo {volume} {10}},\ \bibinfo {pages}
  {022}},\ \Eprint {https://arxiv.org/abs/2405.20566} {arXiv:2405.20566
  [hep-lat]} \BibitemShut {NoStop}%
\bibitem [{\citenamefont {Itou}\ \emph {et~al.}(2025)\citenamefont {Itou},
  \citenamefont {Iida}, \citenamefont {Murakami},\ and\ \citenamefont
  {Suenaga}}]{ItouIida2024}%
  \BibitemOpen
  \bibfield  {author} {\bibinfo {author} {\bibfnamefont {E.}~\bibnamefont
  {Itou}}, \bibinfo {author} {\bibfnamefont {K.}~\bibnamefont {Iida}}, \bibinfo
  {author} {\bibfnamefont {K.}~\bibnamefont {Murakami}},\ and\ \bibinfo
  {author} {\bibfnamefont {D.}~\bibnamefont {Suenaga}},\ }\href
  {https://doi.org/10.22323/1.466.0160} {\bibfield  {journal} {\bibinfo
  {journal} {PoS}\ }\textbf {\bibinfo {volume} {LATTICE2024}},\ \bibinfo
  {pages} {160} (\bibinfo {year} {2025})},\ \Eprint
  {https://arxiv.org/abs/2412.18825} {arXiv:2412.18825 [hep-lat]} \BibitemShut
  {NoStop}%
\bibitem [{\citenamefont {Itou}(2025)}]{ItouReview2025}%
  \BibitemOpen
  \bibfield  {author} {\bibinfo {author} {\bibfnamefont {E.}~\bibnamefont
  {Itou}},\ }\href {https://doi.org/10.3390/universe11110380} {\bibfield
  {journal} {\bibinfo  {journal} {Universe}\ }\textbf {\bibinfo {volume}
  {11}},\ \bibinfo {pages} {380} (\bibinfo {year} {2025})},\ \Eprint
  {https://arxiv.org/abs/2508.03090} {arXiv:2508.03090 [hep-lat]} \BibitemShut
  {NoStop}%
\bibitem [{\citenamefont {Itou}(2026)}]{ItouReview2026}%
  \BibitemOpen
  \bibfield  {author} {\bibinfo {author} {\bibfnamefont {E.}~\bibnamefont
  {Itou}},\ }\href@noop {} {\  (\bibinfo {year} {2026})},\ \Eprint
  {https://arxiv.org/abs/2607.04624} {arXiv:2607.04624 [nucl-th]} \BibitemShut
  {NoStop}%
\bibitem [{\citenamefont {Iida}\ \emph {et~al.}(2026)\citenamefont {Iida},
  \citenamefont {Itou}, \citenamefont {Murakami},\ and\ \citenamefont
  {Suenaga}}]{IidaItou2026}%
  \BibitemOpen
  \bibfield  {author} {\bibinfo {author} {\bibfnamefont {K.}~\bibnamefont
  {Iida}}, \bibinfo {author} {\bibfnamefont {E.}~\bibnamefont {Itou}}, \bibinfo
  {author} {\bibfnamefont {K.}~\bibnamefont {Murakami}},\ and\ \bibinfo
  {author} {\bibfnamefont {D.}~\bibnamefont {Suenaga}},\ }\href@noop {} {\
  (\bibinfo {year} {2026})},\ \Eprint {https://arxiv.org/abs/2606.13974}
  {arXiv:2606.13974 [hep-lat]} \BibitemShut {NoStop}%
\bibitem [{\citenamefont {Hietanen}\ \emph {et~al.}(2014)\citenamefont
  {Hietanen}, \citenamefont {Lewis}, \citenamefont {Pica},\ and\ \citenamefont
  {Sannino}}]{Hietanen2014}%
  \BibitemOpen
  \bibfield  {author} {\bibinfo {author} {\bibfnamefont {A.}~\bibnamefont
  {Hietanen}}, \bibinfo {author} {\bibfnamefont {R.}~\bibnamefont {Lewis}},
  \bibinfo {author} {\bibfnamefont {C.}~\bibnamefont {Pica}},\ and\ \bibinfo
  {author} {\bibfnamefont {F.}~\bibnamefont {Sannino}},\ }\href
  {https://doi.org/10.1007/JHEP12(2014)130} {\bibfield  {journal} {\bibinfo
  {journal} {JHEP}\ }\textbf {\bibinfo {volume} {12}},\ \bibinfo {pages}
  {130}},\ \Eprint {https://arxiv.org/abs/1308.4130} {arXiv:1308.4130 [hep-ph]}
  \BibitemShut {NoStop}%
\bibitem [{\citenamefont {Hochberg}\ \emph {et~al.}(2014)\citenamefont
  {Hochberg}, \citenamefont {Kuflik}, \citenamefont {Volansky},\ and\
  \citenamefont {Wacker}}]{Hochberg2014}%
  \BibitemOpen
  \bibfield  {author} {\bibinfo {author} {\bibfnamefont {Y.}~\bibnamefont
  {Hochberg}}, \bibinfo {author} {\bibfnamefont {E.}~\bibnamefont {Kuflik}},
  \bibinfo {author} {\bibfnamefont {T.}~\bibnamefont {Volansky}},\ and\
  \bibinfo {author} {\bibfnamefont {J.~G.}\ \bibnamefont {Wacker}},\ }\href
  {https://doi.org/10.1103/PhysRevLett.113.171301} {\bibfield  {journal}
  {\bibinfo  {journal} {Phys. Rev. Lett.}\ }\textbf {\bibinfo {volume} {113}},\
  \bibinfo {pages} {171301} (\bibinfo {year} {2014})},\ \Eprint
  {https://arxiv.org/abs/1402.5143} {arXiv:1402.5143 [hep-ph]} \BibitemShut
  {NoStop}%
\bibitem [{\citenamefont {Hochberg}\ \emph {et~al.}(2015)\citenamefont
  {Hochberg}, \citenamefont {Kuflik}, \citenamefont {Murayama}, \citenamefont
  {Volansky},\ and\ \citenamefont {Wacker}}]{Hochberg2015}%
  \BibitemOpen
  \bibfield  {author} {\bibinfo {author} {\bibfnamefont {Y.}~\bibnamefont
  {Hochberg}}, \bibinfo {author} {\bibfnamefont {E.}~\bibnamefont {Kuflik}},
  \bibinfo {author} {\bibfnamefont {H.}~\bibnamefont {Murayama}}, \bibinfo
  {author} {\bibfnamefont {T.}~\bibnamefont {Volansky}},\ and\ \bibinfo
  {author} {\bibfnamefont {J.~G.}\ \bibnamefont {Wacker}},\ }\href
  {https://doi.org/10.1103/PhysRevLett.115.021301} {\bibfield  {journal}
  {\bibinfo  {journal} {Phys. Rev. Lett.}\ }\textbf {\bibinfo {volume} {115}},\
  \bibinfo {pages} {021301} (\bibinfo {year} {2015})},\ \Eprint
  {https://arxiv.org/abs/1411.3727} {arXiv:1411.3727 [hep-ph]} \BibitemShut
  {NoStop}%
\bibitem [{\citenamefont {Bennett}\ \emph {et~al.}(2018)\citenamefont
  {Bennett}, \citenamefont {Hong}, \citenamefont {Lee}, \citenamefont {Lin},
  \citenamefont {Lucini}, \citenamefont {Piai},\ and\ \citenamefont
  {Vadacchino}}]{Bennett2018}%
  \BibitemOpen
  \bibfield  {author} {\bibinfo {author} {\bibfnamefont {E.}~\bibnamefont
  {Bennett}}, \bibinfo {author} {\bibfnamefont {D.~K.}\ \bibnamefont {Hong}},
  \bibinfo {author} {\bibfnamefont {J.-W.}\ \bibnamefont {Lee}}, \bibinfo
  {author} {\bibfnamefont {C.~J.~D.}\ \bibnamefont {Lin}}, \bibinfo {author}
  {\bibfnamefont {B.}~\bibnamefont {Lucini}}, \bibinfo {author} {\bibfnamefont
  {M.}~\bibnamefont {Piai}},\ and\ \bibinfo {author} {\bibfnamefont
  {D.}~\bibnamefont {Vadacchino}},\ }\href
  {https://doi.org/10.1007/JHEP03(2018)185} {\bibfield  {journal} {\bibinfo
  {journal} {JHEP}\ }\textbf {\bibinfo {volume} {03}},\ \bibinfo {pages}
  {185}},\ \Eprint {https://arxiv.org/abs/1712.04220} {arXiv:1712.04220
  [hep-lat]} \BibitemShut {NoStop}%
\bibitem [{\citenamefont {Bennett}\ \emph {et~al.}(2019)\citenamefont
  {Bennett}, \citenamefont {Hong}, \citenamefont {Lee}, \citenamefont {Lin},
  \citenamefont {Lucini}, \citenamefont {Piai},\ and\ \citenamefont
  {Vadacchino}}]{Bennett2019}%
  \BibitemOpen
  \bibfield  {author} {\bibinfo {author} {\bibfnamefont {E.}~\bibnamefont
  {Bennett}}, \bibinfo {author} {\bibfnamefont {D.~K.}\ \bibnamefont {Hong}},
  \bibinfo {author} {\bibfnamefont {J.-W.}\ \bibnamefont {Lee}}, \bibinfo
  {author} {\bibfnamefont {C.~J.~D.}\ \bibnamefont {Lin}}, \bibinfo {author}
  {\bibfnamefont {B.}~\bibnamefont {Lucini}}, \bibinfo {author} {\bibfnamefont
  {M.}~\bibnamefont {Piai}},\ and\ \bibinfo {author} {\bibfnamefont
  {D.}~\bibnamefont {Vadacchino}},\ }\href
  {https://doi.org/10.1007/JHEP12(2019)053} {\bibfield  {journal} {\bibinfo
  {journal} {JHEP}\ }\textbf {\bibinfo {volume} {12}},\ \bibinfo {pages}
  {053}},\ \Eprint {https://arxiv.org/abs/1909.12662} {arXiv:1909.12662
  [hep-lat]} \BibitemShut {NoStop}%
\bibitem [{\citenamefont {Bennett}\ \emph {et~al.}(2024)\citenamefont
  {Bennett}, \citenamefont {Hsiao}, \citenamefont {Lee}, \citenamefont
  {Lucini}, \citenamefont {Maas}, \citenamefont {Piai},\ and\ \citenamefont
  {Zierler}}]{BennettSinglet2024}%
  \BibitemOpen
  \bibfield  {author} {\bibinfo {author} {\bibfnamefont {E.}~\bibnamefont
  {Bennett}}, \bibinfo {author} {\bibfnamefont {H.}~\bibnamefont {Hsiao}},
  \bibinfo {author} {\bibfnamefont {J.-W.}\ \bibnamefont {Lee}}, \bibinfo
  {author} {\bibfnamefont {B.}~\bibnamefont {Lucini}}, \bibinfo {author}
  {\bibfnamefont {A.}~\bibnamefont {Maas}}, \bibinfo {author} {\bibfnamefont
  {M.}~\bibnamefont {Piai}},\ and\ \bibinfo {author} {\bibfnamefont
  {F.}~\bibnamefont {Zierler}},\ }\href
  {https://doi.org/10.1103/PhysRevD.109.034504} {\bibfield  {journal} {\bibinfo
   {journal} {Phys. Rev. D}\ }\textbf {\bibinfo {volume} {109}},\ \bibinfo
  {pages} {034504} (\bibinfo {year} {2024})},\ \Eprint
  {https://arxiv.org/abs/2304.07191} {arXiv:2304.07191 [hep-lat]} \BibitemShut
  {NoStop}%
\bibitem [{\citenamefont {Kulkarni}\ \emph {et~al.}(2023)\citenamefont
  {Kulkarni}, \citenamefont {Maas}, \citenamefont {Mee}, \citenamefont
  {Nikolic}, \citenamefont {Pradler},\ and\ \citenamefont
  {Zierler}}]{KulkarniEtAl}%
  \BibitemOpen
  \bibfield  {author} {\bibinfo {author} {\bibfnamefont {S.}~\bibnamefont
  {Kulkarni}}, \bibinfo {author} {\bibfnamefont {A.}~\bibnamefont {Maas}},
  \bibinfo {author} {\bibfnamefont {S.}~\bibnamefont {Mee}}, \bibinfo {author}
  {\bibfnamefont {M.}~\bibnamefont {Nikolic}}, \bibinfo {author} {\bibfnamefont
  {J.}~\bibnamefont {Pradler}},\ and\ \bibinfo {author} {\bibfnamefont
  {F.}~\bibnamefont {Zierler}},\ }\href
  {https://doi.org/10.21468/SciPostPhys.14.3.044} {\bibfield  {journal}
  {\bibinfo  {journal} {SciPost Phys.}\ }\textbf {\bibinfo {volume} {14}},\
  \bibinfo {pages} {044} (\bibinfo {year} {2023})},\ \Eprint
  {https://arxiv.org/abs/2202.05191} {arXiv:2202.05191 [hep-ph]} \BibitemShut
  {NoStop}%
\bibitem [{\citenamefont {Zierler}\ \emph {et~al.}(2022)\citenamefont
  {Zierler}, \citenamefont {Kulkarni}, \citenamefont {Maas}, \citenamefont
  {Mee}, \citenamefont {Nikolic},\ and\ \citenamefont
  {Pradler}}]{ZierlerStrongDM}%
  \BibitemOpen
  \bibfield  {author} {\bibinfo {author} {\bibfnamefont {F.}~\bibnamefont
  {Zierler}}, \bibinfo {author} {\bibfnamefont {S.}~\bibnamefont {Kulkarni}},
  \bibinfo {author} {\bibfnamefont {A.}~\bibnamefont {Maas}}, \bibinfo {author}
  {\bibfnamefont {S.}~\bibnamefont {Mee}}, \bibinfo {author} {\bibfnamefont
  {M.}~\bibnamefont {Nikolic}},\ and\ \bibinfo {author} {\bibfnamefont
  {J.}~\bibnamefont {Pradler}},\ }\href
  {https://doi.org/10.1051/epjconf/202227408014} {\bibfield  {journal}
  {\bibinfo  {journal} {EPJ Web Conf.}\ }\textbf {\bibinfo {volume} {274}},\
  \bibinfo {pages} {08014} (\bibinfo {year} {2022})},\ \Eprint
  {https://arxiv.org/abs/2211.11272} {arXiv:2211.11272 [hep-ph]} \BibitemShut
  {NoStop}%
\bibitem [{\citenamefont {Sonin}(1987)}]{Sonin1987}%
  \BibitemOpen
  \bibfield  {author} {\bibinfo {author} {\bibfnamefont {E.~B.}\ \bibnamefont
  {Sonin}},\ }\href {https://doi.org/10.1103/RevModPhys.59.87} {\bibfield
  {journal} {\bibinfo  {journal} {Rev. Mod. Phys.}\ }\textbf {\bibinfo {volume}
  {59}},\ \bibinfo {pages} {87} (\bibinfo {year} {1987})}\BibitemShut {NoStop}%
\bibitem [{\citenamefont {Salomaa}\ and\ \citenamefont
  {Volovik}(1987)}]{SalomaaVolovik1987}%
  \BibitemOpen
  \bibfield  {author} {\bibinfo {author} {\bibfnamefont {M.~M.}\ \bibnamefont
  {Salomaa}}\ and\ \bibinfo {author} {\bibfnamefont {G.~E.}\ \bibnamefont
  {Volovik}},\ }\href {https://doi.org/10.1103/RevModPhys.59.533} {\bibfield
  {journal} {\bibinfo  {journal} {Rev. Mod. Phys.}\ }\textbf {\bibinfo {volume}
  {59}},\ \bibinfo {pages} {533} (\bibinfo {year} {1987})}\BibitemShut
  {NoStop}%
\bibitem [{\citenamefont {Fetter}(2009)}]{Fetter2009}%
  \BibitemOpen
  \bibfield  {author} {\bibinfo {author} {\bibfnamefont {A.~L.}\ \bibnamefont
  {Fetter}},\ }\href {https://doi.org/10.1103/RevModPhys.81.647} {\bibfield
  {journal} {\bibinfo  {journal} {Rev. Mod. Phys.}\ }\textbf {\bibinfo {volume}
  {81}},\ \bibinfo {pages} {647} (\bibinfo {year} {2009})},\ \Eprint
  {https://arxiv.org/abs/0801.2952} {arXiv:0801.2952 [cond-mat.stat-mech]}
  \BibitemShut {NoStop}%
\bibitem [{\citenamefont {Tsubota}\ and\ \citenamefont
  {Kasamatsu}(2025)}]{TsubotaKasamatsu2025}%
  \BibitemOpen
  \bibfield  {author} {\bibinfo {author} {\bibfnamefont {M.}~\bibnamefont
  {Tsubota}}\ and\ \bibinfo {author} {\bibfnamefont {K.}~\bibnamefont
  {Kasamatsu}},\ }\href {https://doi.org/10.1093/oso/9780198742944.001.0001}
  {\emph {\bibinfo {title} {Quantum Hydrodynamics and Turbulence}}}\ (\bibinfo
  {publisher} {Oxford University Press},\ \bibinfo {year} {2025})\BibitemShut
  {NoStop}%
\bibitem [{\citenamefont {Alford}\ \emph {et~al.}(2008)\citenamefont {Alford},
  \citenamefont {Schmitt}, \citenamefont {Rajagopal},\ and\ \citenamefont
  {Sch{\"a}fer}}]{AlfordReview2008}%
  \BibitemOpen
  \bibfield  {author} {\bibinfo {author} {\bibfnamefont {M.~G.}\ \bibnamefont
  {Alford}}, \bibinfo {author} {\bibfnamefont {A.}~\bibnamefont {Schmitt}},
  \bibinfo {author} {\bibfnamefont {K.}~\bibnamefont {Rajagopal}},\ and\
  \bibinfo {author} {\bibfnamefont {T.}~\bibnamefont {Sch{\"a}fer}},\ }\href
  {https://doi.org/10.1103/RevModPhys.80.1455} {\bibfield  {journal} {\bibinfo
  {journal} {Rev. Mod. Phys.}\ }\textbf {\bibinfo {volume} {80}},\ \bibinfo
  {pages} {1455} (\bibinfo {year} {2008})},\ \Eprint
  {https://arxiv.org/abs/0709.4635} {arXiv:0709.4635 [hep-ph]} \BibitemShut
  {NoStop}%
\bibitem [{\citenamefont {Eto}\ \emph {et~al.}(2014)\citenamefont {Eto},
  \citenamefont {Hirono}, \citenamefont {Nitta},\ and\ \citenamefont
  {Yasui}}]{EtoDenseReview2014}%
  \BibitemOpen
  \bibfield  {author} {\bibinfo {author} {\bibfnamefont {M.}~\bibnamefont
  {Eto}}, \bibinfo {author} {\bibfnamefont {Y.}~\bibnamefont {Hirono}},
  \bibinfo {author} {\bibfnamefont {M.}~\bibnamefont {Nitta}},\ and\ \bibinfo
  {author} {\bibfnamefont {S.}~\bibnamefont {Yasui}},\ }\href
  {https://doi.org/10.1093/ptep/ptt095} {\bibfield  {journal} {\bibinfo
  {journal} {PTEP}\ }\textbf {\bibinfo {volume} {2014}},\ \bibinfo {pages}
  {012D01} (\bibinfo {year} {2014})},\ \Eprint
  {https://arxiv.org/abs/1308.1535} {arXiv:1308.1535 [hep-ph]} \BibitemShut
  {NoStop}%
\bibitem [{\citenamefont {Alford}\ \emph {et~al.}(2019)\citenamefont {Alford},
  \citenamefont {Baym}, \citenamefont {Fukushima}, \citenamefont {Hatsuda},\
  and\ \citenamefont {Tachibana}}]{AlfordBaymFukushima2019}%
  \BibitemOpen
  \bibfield  {author} {\bibinfo {author} {\bibfnamefont {M.~G.}\ \bibnamefont
  {Alford}}, \bibinfo {author} {\bibfnamefont {G.}~\bibnamefont {Baym}},
  \bibinfo {author} {\bibfnamefont {K.}~\bibnamefont {Fukushima}}, \bibinfo
  {author} {\bibfnamefont {T.}~\bibnamefont {Hatsuda}},\ and\ \bibinfo {author}
  {\bibfnamefont {M.}~\bibnamefont {Tachibana}},\ }\href
  {https://doi.org/10.1103/PhysRevD.99.036004} {\bibfield  {journal} {\bibinfo
  {journal} {Phys. Rev. D}\ }\textbf {\bibinfo {volume} {99}},\ \bibinfo
  {pages} {036004} (\bibinfo {year} {2019})},\ \Eprint
  {https://arxiv.org/abs/1803.05115} {arXiv:1803.05115 [hep-ph]} \BibitemShut
  {NoStop}%
\bibitem [{\citenamefont {Wess}\ and\ \citenamefont
  {Zumino}(1971)}]{WessZumino1971}%
  \BibitemOpen
  \bibfield  {author} {\bibinfo {author} {\bibfnamefont {J.}~\bibnamefont
  {Wess}}\ and\ \bibinfo {author} {\bibfnamefont {B.}~\bibnamefont {Zumino}},\
  }\href {https://doi.org/10.1016/0370-2693(71)90582-X} {\bibfield  {journal}
  {\bibinfo  {journal} {Phys. Lett. B}\ }\textbf {\bibinfo {volume} {37}},\
  \bibinfo {pages} {95} (\bibinfo {year} {1971})}\BibitemShut {NoStop}%
\bibitem [{\citenamefont {Witten}(1983)}]{Witten1983}%
  \BibitemOpen
  \bibfield  {author} {\bibinfo {author} {\bibfnamefont {E.}~\bibnamefont
  {Witten}},\ }\href {https://doi.org/10.1016/0550-3213(83)90063-9} {\bibfield
  {journal} {\bibinfo  {journal} {Nucl. Phys. B}\ }\textbf {\bibinfo {volume}
  {223}},\ \bibinfo {pages} {422} (\bibinfo {year} {1983})}\BibitemShut
  {NoStop}%
\bibitem [{\citenamefont {Callan}\ and\ \citenamefont
  {Harvey}(1985)}]{CallanHarvey1985}%
  \BibitemOpen
  \bibfield  {author} {\bibinfo {author} {\bibfnamefont {C.~G.}\ \bibnamefont
  {Callan}, \bibfnamefont {Jr.}}\ and\ \bibinfo {author} {\bibfnamefont
  {J.~A.}\ \bibnamefont {Harvey}},\ }\href
  {https://doi.org/10.1016/0550-3213(85)90489-4} {\bibfield  {journal}
  {\bibinfo  {journal} {Nucl. Phys. B}\ }\textbf {\bibinfo {volume} {250}},\
  \bibinfo {pages} {427} (\bibinfo {year} {1985})}\BibitemShut {NoStop}%
\bibitem [{\citenamefont {Fukushima}\ and\ \citenamefont
  {Imaki}(2018)}]{FukushimaImaki2018}%
  \BibitemOpen
  \bibfield  {author} {\bibinfo {author} {\bibfnamefont {K.}~\bibnamefont
  {Fukushima}}\ and\ \bibinfo {author} {\bibfnamefont {S.}~\bibnamefont
  {Imaki}},\ }\href {https://doi.org/10.1103/PhysRevD.97.114003} {\bibfield
  {journal} {\bibinfo  {journal} {Phys. Rev. D}\ }\textbf {\bibinfo {volume}
  {97}},\ \bibinfo {pages} {114003} (\bibinfo {year} {2018})},\ \Eprint
  {https://arxiv.org/abs/1802.08096} {arXiv:1802.08096 [hep-ph]} \BibitemShut
  {NoStop}%
\bibitem [{\citenamefont {Witten}(1984)}]{WittenBosonization1984}%
  \BibitemOpen
  \bibfield  {author} {\bibinfo {author} {\bibfnamefont {E.}~\bibnamefont
  {Witten}},\ }\href {https://doi.org/10.1007/BF01215276} {\bibfield  {journal}
  {\bibinfo  {journal} {Commun. Math. Phys.}\ }\textbf {\bibinfo {volume}
  {92}},\ \bibinfo {pages} {455} (\bibinfo {year} {1984})}\BibitemShut
  {NoStop}%
\bibitem [{\citenamefont {Knizhnik}\ and\ \citenamefont
  {Zamolodchikov}(1984)}]{KZ1984}%
  \BibitemOpen
  \bibfield  {author} {\bibinfo {author} {\bibfnamefont {V.~G.}\ \bibnamefont
  {Knizhnik}}\ and\ \bibinfo {author} {\bibfnamefont {A.~B.}\ \bibnamefont
  {Zamolodchikov}},\ }\href {https://doi.org/10.1016/0550-3213(84)90374-2}
  {\bibfield  {journal} {\bibinfo  {journal} {Nucl. Phys. B}\ }\textbf
  {\bibinfo {volume} {247}},\ \bibinfo {pages} {83} (\bibinfo {year}
  {1984})}\BibitemShut {NoStop}%
\bibitem [{\citenamefont {Belavin}\ \emph {et~al.}(1984)\citenamefont
  {Belavin}, \citenamefont {Polyakov},\ and\ \citenamefont
  {Zamolodchikov}}]{BPZ1984}%
  \BibitemOpen
  \bibfield  {author} {\bibinfo {author} {\bibfnamefont {A.~A.}\ \bibnamefont
  {Belavin}}, \bibinfo {author} {\bibfnamefont {A.~M.}\ \bibnamefont
  {Polyakov}},\ and\ \bibinfo {author} {\bibfnamefont {A.~B.}\ \bibnamefont
  {Zamolodchikov}},\ }\href {https://doi.org/10.1016/0550-3213(84)90052-X}
  {\bibfield  {journal} {\bibinfo  {journal} {Nucl. Phys. B}\ }\textbf
  {\bibinfo {volume} {241}},\ \bibinfo {pages} {333} (\bibinfo {year}
  {1984})}\BibitemShut {NoStop}%
\bibitem [{\citenamefont {Di~Francesco}\ \emph {et~al.}(1997)\citenamefont
  {Di~Francesco}, \citenamefont {Mathieu},\ and\ \citenamefont
  {Senechal}}]{DiFrancesco1997}%
  \BibitemOpen
  \bibfield  {author} {\bibinfo {author} {\bibfnamefont {P.}~\bibnamefont
  {Di~Francesco}}, \bibinfo {author} {\bibfnamefont {P.}~\bibnamefont
  {Mathieu}},\ and\ \bibinfo {author} {\bibfnamefont {D.}~\bibnamefont
  {Senechal}},\ }\href {https://doi.org/10.1007/978-1-4612-2256-9} {\emph
  {\bibinfo {title} {{Conformal Field Theory}}}},\ Graduate Texts in
  Contemporary Physics\ (\bibinfo  {publisher} {Springer-Verlag},\ \bibinfo
  {address} {New York},\ \bibinfo {year} {1997})\BibitemShut {NoStop}%
\bibitem [{\citenamefont {Kanazawa}\ \emph {et~al.}(2009)\citenamefont
  {Kanazawa}, \citenamefont {Wettig},\ and\ \citenamefont
  {Yamamoto}}]{KanazawaWettigYamamoto2009}%
  \BibitemOpen
  \bibfield  {author} {\bibinfo {author} {\bibfnamefont {T.}~\bibnamefont
  {Kanazawa}}, \bibinfo {author} {\bibfnamefont {T.}~\bibnamefont {Wettig}},\
  and\ \bibinfo {author} {\bibfnamefont {N.}~\bibnamefont {Yamamoto}},\ }\href
  {https://doi.org/10.1088/1126-6708/2009/08/003} {\bibfield  {journal}
  {\bibinfo  {journal} {JHEP}\ }\textbf {\bibinfo {volume} {08}},\ \bibinfo
  {pages} {003}},\ \Eprint {https://arxiv.org/abs/0906.3579} {arXiv:0906.3579
  [hep-ph]} \BibitemShut {NoStop}%
\bibitem [{\citenamefont {Brauner}\ and\ \citenamefont
  {Kole{\v{s}}ov{\'a}}(2019)}]{BraunerKolesova2019}%
  \BibitemOpen
  \bibfield  {author} {\bibinfo {author} {\bibfnamefont {T.}~\bibnamefont
  {Brauner}}\ and\ \bibinfo {author} {\bibfnamefont {H.}~\bibnamefont
  {Kole{\v{s}}ov{\'a}}},\ }\href
  {https://doi.org/10.1016/j.nuclphysb.2019.114676} {\bibfield  {journal}
  {\bibinfo  {journal} {Nucl. Phys. B}\ }\textbf {\bibinfo {volume} {945}},\
  \bibinfo {pages} {114676} (\bibinfo {year} {2019})},\ \Eprint
  {https://arxiv.org/abs/1809.05310} {arXiv:1809.05310 [hep-th]} \BibitemShut
  {NoStop}%
\bibitem [{\citenamefont {Lee}\ \emph {et~al.}(2021)\citenamefont {Lee},
  \citenamefont {Ohmori},\ and\ \citenamefont
  {Tachikawa}}]{LeeOhmoriTachikawa}%
  \BibitemOpen
  \bibfield  {author} {\bibinfo {author} {\bibfnamefont {Y.}~\bibnamefont
  {Lee}}, \bibinfo {author} {\bibfnamefont {K.}~\bibnamefont {Ohmori}},\ and\
  \bibinfo {author} {\bibfnamefont {Y.}~\bibnamefont {Tachikawa}},\ }\href
  {https://doi.org/10.21468/SciPostPhys.10.3.061} {\bibfield  {journal}
  {\bibinfo  {journal} {SciPost Phys.}\ }\textbf {\bibinfo {volume} {10}},\
  \bibinfo {pages} {061} (\bibinfo {year} {2021})},\ \Eprint
  {https://arxiv.org/abs/2009.00033} {arXiv:2009.00033 [hep-th]} \BibitemShut
  {NoStop}%
\bibitem [{\citenamefont {Saito}(2024)}]{Saito}%
  \BibitemOpen
  \bibfield  {author} {\bibinfo {author} {\bibfnamefont {S.}~\bibnamefont
  {Saito}},\ }\href {https://doi.org/10.1007/JHEP10(2024)099} {\bibfield
  {journal} {\bibinfo  {journal} {JHEP}\ }\textbf {\bibinfo {volume} {10}},\
  \bibinfo {pages} {099}},\ \Eprint {https://arxiv.org/abs/2404.06185}
  {arXiv:2404.06185 [hep-th]} \BibitemShut {NoStop}%
\bibitem [{\citenamefont {Manton}(1982)}]{Manton1982}%
  \BibitemOpen
  \bibfield  {author} {\bibinfo {author} {\bibfnamefont {N.~S.}\ \bibnamefont
  {Manton}},\ }\href {https://doi.org/10.1016/0370-2693(82)90950-9} {\bibfield
  {journal} {\bibinfo  {journal} {Phys. Lett. B}\ }\textbf {\bibinfo {volume}
  {110}},\ \bibinfo {pages} {54} (\bibinfo {year} {1982})}\BibitemShut
  {NoStop}%
\bibitem [{\citenamefont {Eto}\ \emph {et~al.}(2006{\natexlab{a}})\citenamefont
  {Eto}, \citenamefont {Isozumi}, \citenamefont {Nitta}, \citenamefont
  {Ohashi},\ and\ \citenamefont {Sakai}}]{EtoEffective2006}%
  \BibitemOpen
  \bibfield  {author} {\bibinfo {author} {\bibfnamefont {M.}~\bibnamefont
  {Eto}}, \bibinfo {author} {\bibfnamefont {Y.}~\bibnamefont {Isozumi}},
  \bibinfo {author} {\bibfnamefont {M.}~\bibnamefont {Nitta}}, \bibinfo
  {author} {\bibfnamefont {K.}~\bibnamefont {Ohashi}},\ and\ \bibinfo {author}
  {\bibfnamefont {N.}~\bibnamefont {Sakai}},\ }\href
  {https://doi.org/10.1103/PhysRevD.73.125008} {\bibfield  {journal} {\bibinfo
  {journal} {Phys. Rev. D}\ }\textbf {\bibinfo {volume} {73}},\ \bibinfo
  {pages} {125008} (\bibinfo {year} {2006}{\natexlab{a}})},\ \Eprint
  {https://arxiv.org/abs/hep-th/0602289} {arXiv:hep-th/0602289} \BibitemShut
  {NoStop}%
\bibitem [{\citenamefont {Qiu}\ and\ \citenamefont
  {Nitta}(2024)}]{QiuNitta2024}%
  \BibitemOpen
  \bibfield  {author} {\bibinfo {author} {\bibfnamefont {Z.}~\bibnamefont
  {Qiu}}\ and\ \bibinfo {author} {\bibfnamefont {M.}~\bibnamefont {Nitta}},\
  }\href {https://doi.org/10.1007/JHEP06(2024)139} {\bibfield  {journal}
  {\bibinfo  {journal} {JHEP}\ }\textbf {\bibinfo {volume} {06}},\ \bibinfo
  {pages} {139}},\ \Eprint {https://arxiv.org/abs/2403.07433} {arXiv:2403.07433
  [hep-ph]} \BibitemShut {NoStop}%
\bibitem [{\citenamefont {Hamada}\ \emph
  {et~al.}(2026{\natexlab{a}})\citenamefont {Hamada}, \citenamefont {Nitta},\
  and\ \citenamefont {Qiu}}]{QiuNitta2025}%
  \BibitemOpen
  \bibfield  {author} {\bibinfo {author} {\bibfnamefont {Y.}~\bibnamefont
  {Hamada}}, \bibinfo {author} {\bibfnamefont {M.}~\bibnamefont {Nitta}},\ and\
  \bibinfo {author} {\bibfnamefont {Z.}~\bibnamefont {Qiu}},\ }\href
  {https://doi.org/10.1007/JHEP02(2026)200} {\bibfield  {journal} {\bibinfo
  {journal} {JHEP}\ }\textbf {\bibinfo {volume} {02}},\ \bibinfo {pages}
  {200}},\ \Eprint {https://arxiv.org/abs/2509.20844} {arXiv:2509.20844
  [hep-ph]} \BibitemShut {NoStop}%
\bibitem [{\citenamefont {Hamada}\ \emph
  {et~al.}(2026{\natexlab{b}})\citenamefont {Hamada}, \citenamefont {Nitta},\
  and\ \citenamefont {Qiu}}]{QiuNitta2026}%
  \BibitemOpen
  \bibfield  {author} {\bibinfo {author} {\bibfnamefont {Y.}~\bibnamefont
  {Hamada}}, \bibinfo {author} {\bibfnamefont {M.}~\bibnamefont {Nitta}},\ and\
  \bibinfo {author} {\bibfnamefont {Z.}~\bibnamefont {Qiu}},\ }\href@noop {} {\
   (\bibinfo {year} {2026}{\natexlab{b}})},\ \Eprint
  {https://arxiv.org/abs/2602.11762} {arXiv:2602.11762 [hep-ph]} \BibitemShut
  {NoStop}%
\bibitem [{\citenamefont {Polyakov}\ and\ \citenamefont
  {Wiegmann}(1984)}]{PolyakovWiegmann1984}%
  \BibitemOpen
  \bibfield  {author} {\bibinfo {author} {\bibfnamefont {A.~M.}\ \bibnamefont
  {Polyakov}}\ and\ \bibinfo {author} {\bibfnamefont {P.~B.}\ \bibnamefont
  {Wiegmann}},\ }\href {https://doi.org/10.1016/0370-2693(84)90206-5}
  {\bibfield  {journal} {\bibinfo  {journal} {Phys. Lett. B}\ }\textbf
  {\bibinfo {volume} {141}},\ \bibinfo {pages} {223} (\bibinfo {year}
  {1984})}\BibitemShut {NoStop}%
\bibitem [{\citenamefont {Schubring}\ and\ \citenamefont
  {Shifman}(2021)}]{SchubringShifman2020}%
  \BibitemOpen
  \bibfield  {author} {\bibinfo {author} {\bibfnamefont {D.}~\bibnamefont
  {Schubring}}\ and\ \bibinfo {author} {\bibfnamefont {M.}~\bibnamefont
  {Shifman}},\ }\href {https://doi.org/10.1103/PhysRevD.103.025016} {\bibfield
  {journal} {\bibinfo  {journal} {Phys. Rev. D}\ }\textbf {\bibinfo {volume}
  {103}},\ \bibinfo {pages} {025016} (\bibinfo {year} {2021})},\ \Eprint
  {https://arxiv.org/abs/2002.04696} {arXiv:2002.04696 [hep-th]} \BibitemShut
  {NoStop}%
\bibitem [{\citenamefont {Coleman}(1973)}]{Coleman1973}%
  \BibitemOpen
  \bibfield  {author} {\bibinfo {author} {\bibfnamefont {S.~R.}\ \bibnamefont
  {Coleman}},\ }\href {https://doi.org/10.1007/BF01646487} {\bibfield
  {journal} {\bibinfo  {journal} {Commun. Math. Phys.}\ }\textbf {\bibinfo
  {volume} {31}},\ \bibinfo {pages} {259} (\bibinfo {year} {1973})}\BibitemShut
  {NoStop}%
\bibitem [{\citenamefont {Hanany}\ and\ \citenamefont
  {Tong}(2003)}]{HananyTong2003}%
  \BibitemOpen
  \bibfield  {author} {\bibinfo {author} {\bibfnamefont {A.}~\bibnamefont
  {Hanany}}\ and\ \bibinfo {author} {\bibfnamefont {D.}~\bibnamefont {Tong}},\
  }\href {https://doi.org/10.1088/1126-6708/2003/07/037} {\bibfield  {journal}
  {\bibinfo  {journal} {JHEP}\ }\textbf {\bibinfo {volume} {07}},\ \bibinfo
  {pages} {037}},\ \Eprint {https://arxiv.org/abs/hep-th/0306150}
  {arXiv:hep-th/0306150} \BibitemShut {NoStop}%
\bibitem [{\citenamefont {Auzzi}\ \emph {et~al.}(2003)\citenamefont {Auzzi},
  \citenamefont {Bolognesi}, \citenamefont {Evslin}, \citenamefont {Konishi},\
  and\ \citenamefont {Yung}}]{AuzziEtAl2003}%
  \BibitemOpen
  \bibfield  {author} {\bibinfo {author} {\bibfnamefont {R.}~\bibnamefont
  {Auzzi}}, \bibinfo {author} {\bibfnamefont {S.}~\bibnamefont {Bolognesi}},
  \bibinfo {author} {\bibfnamefont {J.}~\bibnamefont {Evslin}}, \bibinfo
  {author} {\bibfnamefont {K.}~\bibnamefont {Konishi}},\ and\ \bibinfo {author}
  {\bibfnamefont {A.}~\bibnamefont {Yung}},\ }\href
  {https://doi.org/10.1016/j.nuclphysb.2003.09.029} {\bibfield  {journal}
  {\bibinfo  {journal} {Nucl. Phys. B}\ }\textbf {\bibinfo {volume} {673}},\
  \bibinfo {pages} {187} (\bibinfo {year} {2003})},\ \Eprint
  {https://arxiv.org/abs/hep-th/0307287} {arXiv:hep-th/0307287} \BibitemShut
  {NoStop}%
\bibitem [{\citenamefont {Eto}\ \emph {et~al.}(2006{\natexlab{b}})\citenamefont
  {Eto}, \citenamefont {Isozumi}, \citenamefont {Nitta}, \citenamefont
  {Ohashi},\ and\ \citenamefont {Sakai}}]{EtoEtAl2006}%
  \BibitemOpen
  \bibfield  {author} {\bibinfo {author} {\bibfnamefont {M.}~\bibnamefont
  {Eto}}, \bibinfo {author} {\bibfnamefont {Y.}~\bibnamefont {Isozumi}},
  \bibinfo {author} {\bibfnamefont {M.}~\bibnamefont {Nitta}}, \bibinfo
  {author} {\bibfnamefont {K.}~\bibnamefont {Ohashi}},\ and\ \bibinfo {author}
  {\bibfnamefont {N.}~\bibnamefont {Sakai}},\ }\href
  {https://doi.org/10.1103/PhysRevLett.96.161601} {\bibfield  {journal}
  {\bibinfo  {journal} {Phys. Rev. Lett.}\ }\textbf {\bibinfo {volume} {96}},\
  \bibinfo {pages} {161601} (\bibinfo {year} {2006}{\natexlab{b}})},\ \Eprint
  {https://arxiv.org/abs/hep-th/0511088} {arXiv:hep-th/0511088} \BibitemShut
  {NoStop}%
\bibitem [{\citenamefont {Eto}\ \emph {et~al.}(2006{\natexlab{c}})\citenamefont
  {Eto}, \citenamefont {Konishi}, \citenamefont {Marmorini}, \citenamefont
  {Nitta}, \citenamefont {Ohashi}, \citenamefont {Vinci},\ and\ \citenamefont
  {Yokoi}}]{EtoHigherWinding2006}%
  \BibitemOpen
  \bibfield  {author} {\bibinfo {author} {\bibfnamefont {M.}~\bibnamefont
  {Eto}}, \bibinfo {author} {\bibfnamefont {K.}~\bibnamefont {Konishi}},
  \bibinfo {author} {\bibfnamefont {G.}~\bibnamefont {Marmorini}}, \bibinfo
  {author} {\bibfnamefont {M.}~\bibnamefont {Nitta}}, \bibinfo {author}
  {\bibfnamefont {K.}~\bibnamefont {Ohashi}}, \bibinfo {author} {\bibfnamefont
  {W.}~\bibnamefont {Vinci}},\ and\ \bibinfo {author} {\bibfnamefont
  {N.}~\bibnamefont {Yokoi}},\ }\href
  {https://doi.org/10.1103/PhysRevD.74.065021} {\bibfield  {journal} {\bibinfo
  {journal} {Phys. Rev. D}\ }\textbf {\bibinfo {volume} {74}},\ \bibinfo
  {pages} {065021} (\bibinfo {year} {2006}{\natexlab{c}})},\ \Eprint
  {https://arxiv.org/abs/hep-th/0607070} {arXiv:hep-th/0607070} \BibitemShut
  {NoStop}%
\bibitem [{\citenamefont {Tong}(2005)}]{TongTASI2005}%
  \BibitemOpen
  \bibfield  {author} {\bibinfo {author} {\bibfnamefont {D.}~\bibnamefont
  {Tong}},\ }in\ \href@noop {} {\emph {\bibinfo {booktitle} {{Theoretical
  Advanced Study Institute in Elementary Particle Physics}: {Many Dimensions of
  String Theory}}}}\ (\bibinfo {year} {2005})\ \Eprint
  {https://arxiv.org/abs/hep-th/0509216} {arXiv:hep-th/0509216} \BibitemShut
  {NoStop}%
\bibitem [{\citenamefont {Eto}\ \emph {et~al.}(2006{\natexlab{d}})\citenamefont
  {Eto}, \citenamefont {Isozumi}, \citenamefont {Nitta}, \citenamefont
  {Ohashi},\ and\ \citenamefont {Sakai}}]{EtoReview2006}%
  \BibitemOpen
  \bibfield  {author} {\bibinfo {author} {\bibfnamefont {M.}~\bibnamefont
  {Eto}}, \bibinfo {author} {\bibfnamefont {Y.}~\bibnamefont {Isozumi}},
  \bibinfo {author} {\bibfnamefont {M.}~\bibnamefont {Nitta}}, \bibinfo
  {author} {\bibfnamefont {K.}~\bibnamefont {Ohashi}},\ and\ \bibinfo {author}
  {\bibfnamefont {N.}~\bibnamefont {Sakai}},\ }\href
  {https://doi.org/10.1088/0305-4470/39/26/R01} {\bibfield  {journal} {\bibinfo
   {journal} {J. Phys. A}\ }\textbf {\bibinfo {volume} {39}},\ \bibinfo {pages}
  {R315} (\bibinfo {year} {2006}{\natexlab{d}})},\ \Eprint
  {https://arxiv.org/abs/hep-th/0602170} {arXiv:hep-th/0602170} \BibitemShut
  {NoStop}%
\bibitem [{\citenamefont {Shifman}\ and\ \citenamefont
  {Yung}(2007)}]{ShifmanYungRMP2007}%
  \BibitemOpen
  \bibfield  {author} {\bibinfo {author} {\bibfnamefont {M.}~\bibnamefont
  {Shifman}}\ and\ \bibinfo {author} {\bibfnamefont {A.}~\bibnamefont {Yung}},\
  }\href {https://doi.org/10.1103/RevModPhys.79.1139} {\bibfield  {journal}
  {\bibinfo  {journal} {Rev. Mod. Phys.}\ }\textbf {\bibinfo {volume} {79}},\
  \bibinfo {pages} {1139} (\bibinfo {year} {2007})},\ \Eprint
  {https://arxiv.org/abs/hep-th/0703267} {arXiv:hep-th/0703267} \BibitemShut
  {NoStop}%
\bibitem [{\citenamefont {Shifman}\ and\ \citenamefont
  {Yung}(2009)}]{ShifmanYungBook2009}%
  \BibitemOpen
  \bibfield  {author} {\bibinfo {author} {\bibfnamefont {M.}~\bibnamefont
  {Shifman}}\ and\ \bibinfo {author} {\bibfnamefont {A.}~\bibnamefont {Yung}},\
  }\href {https://doi.org/10.1017/9781009402200} {\emph {\bibinfo {title}
  {{Supersymmetric Solitons}}}}\ (\bibinfo  {publisher} {Cambridge University
  Press},\ \bibinfo {year} {2009})\BibitemShut {NoStop}%
\bibitem [{\citenamefont {Tong}(2009)}]{TongReview2009}%
  \BibitemOpen
  \bibfield  {author} {\bibinfo {author} {\bibfnamefont {D.}~\bibnamefont
  {Tong}},\ }\href {https://doi.org/10.1016/j.aop.2008.10.005} {\bibfield
  {journal} {\bibinfo  {journal} {Annals Phys.}\ }\textbf {\bibinfo {volume}
  {324}},\ \bibinfo {pages} {30} (\bibinfo {year} {2009})},\ \Eprint
  {https://arxiv.org/abs/0809.5060} {arXiv:0809.5060 [hep-th]} \BibitemShut
  {NoStop}%
\bibitem [{\citenamefont {Shifman}\ and\ \citenamefont
  {Yung}(2004)}]{ShifmanYungMonopoles2004}%
  \BibitemOpen
  \bibfield  {author} {\bibinfo {author} {\bibfnamefont {M.}~\bibnamefont
  {Shifman}}\ and\ \bibinfo {author} {\bibfnamefont {A.}~\bibnamefont {Yung}},\
  }\href {https://doi.org/10.1103/PhysRevD.70.045004} {\bibfield  {journal}
  {\bibinfo  {journal} {Phys. Rev. D}\ }\textbf {\bibinfo {volume} {70}},\
  \bibinfo {pages} {045004} (\bibinfo {year} {2004})},\ \Eprint
  {https://arxiv.org/abs/hep-th/0403149} {arXiv:hep-th/0403149} \BibitemShut
  {NoStop}%
\bibitem [{\citenamefont {Hanany}\ and\ \citenamefont
  {Tong}(2004)}]{HananyTongQuantum2004}%
  \BibitemOpen
  \bibfield  {author} {\bibinfo {author} {\bibfnamefont {A.}~\bibnamefont
  {Hanany}}\ and\ \bibinfo {author} {\bibfnamefont {D.}~\bibnamefont {Tong}},\
  }\href {https://doi.org/10.1088/1126-6708/2004/04/066} {\bibfield  {journal}
  {\bibinfo  {journal} {JHEP}\ }\textbf {\bibinfo {volume} {04}},\ \bibinfo
  {pages} {066}},\ \Eprint {https://arxiv.org/abs/hep-th/0403158}
  {arXiv:hep-th/0403158} \BibitemShut {NoStop}%
\bibitem [{\citenamefont {Balachandran}\ \emph {et~al.}(2006)\citenamefont
  {Balachandran}, \citenamefont {Digal},\ and\ \citenamefont
  {Matsuura}}]{BalachandranDigalMatsuura2006}%
  \BibitemOpen
  \bibfield  {author} {\bibinfo {author} {\bibfnamefont {A.~P.}\ \bibnamefont
  {Balachandran}}, \bibinfo {author} {\bibfnamefont {S.}~\bibnamefont
  {Digal}},\ and\ \bibinfo {author} {\bibfnamefont {T.}~\bibnamefont
  {Matsuura}},\ }\href {https://doi.org/10.1103/PhysRevD.73.074009} {\bibfield
  {journal} {\bibinfo  {journal} {Phys. Rev. D}\ }\textbf {\bibinfo {volume}
  {73}},\ \bibinfo {pages} {074009} (\bibinfo {year} {2006})},\ \Eprint
  {https://arxiv.org/abs/hep-ph/0509276} {arXiv:hep-ph/0509276} \BibitemShut
  {NoStop}%
\bibitem [{\citenamefont {Eto}\ and\ \citenamefont
  {Nitta}(2009)}]{EtoNitta2009}%
  \BibitemOpen
  \bibfield  {author} {\bibinfo {author} {\bibfnamefont {M.}~\bibnamefont
  {Eto}}\ and\ \bibinfo {author} {\bibfnamefont {M.}~\bibnamefont {Nitta}},\
  }\href {https://doi.org/10.1103/PhysRevD.80.125007} {\bibfield  {journal}
  {\bibinfo  {journal} {Phys. Rev. D}\ }\textbf {\bibinfo {volume} {80}},\
  \bibinfo {pages} {125007} (\bibinfo {year} {2009})},\ \Eprint
  {https://arxiv.org/abs/0907.1278} {arXiv:0907.1278 [hep-ph]} \BibitemShut
  {NoStop}%
\bibitem [{\citenamefont {Nakano}\ \emph {et~al.}(2008)\citenamefont {Nakano},
  \citenamefont {Nitta},\ and\ \citenamefont
  {Matsuura}}]{NakanoNittaMatsuura2008}%
  \BibitemOpen
  \bibfield  {author} {\bibinfo {author} {\bibfnamefont {E.}~\bibnamefont
  {Nakano}}, \bibinfo {author} {\bibfnamefont {M.}~\bibnamefont {Nitta}},\ and\
  \bibinfo {author} {\bibfnamefont {T.}~\bibnamefont {Matsuura}},\ }\href
  {https://doi.org/10.1103/PhysRevD.78.045002} {\bibfield  {journal} {\bibinfo
  {journal} {Phys. Rev. D}\ }\textbf {\bibinfo {volume} {78}},\ \bibinfo
  {pages} {045002} (\bibinfo {year} {2008})},\ \Eprint
  {https://arxiv.org/abs/0708.4096} {arXiv:0708.4096 [hep-ph]} \BibitemShut
  {NoStop}%
\bibitem [{\citenamefont {Eto}\ \emph {et~al.}(2009)\citenamefont {Eto},
  \citenamefont {Nakano},\ and\ \citenamefont {Nitta}}]{EtoNakanoNitta2009}%
  \BibitemOpen
  \bibfield  {author} {\bibinfo {author} {\bibfnamefont {M.}~\bibnamefont
  {Eto}}, \bibinfo {author} {\bibfnamefont {E.}~\bibnamefont {Nakano}},\ and\
  \bibinfo {author} {\bibfnamefont {M.}~\bibnamefont {Nitta}},\ }\href
  {https://doi.org/10.1103/PhysRevD.80.125011} {\bibfield  {journal} {\bibinfo
  {journal} {Phys. Rev. D}\ }\textbf {\bibinfo {volume} {80}},\ \bibinfo
  {pages} {125011} (\bibinfo {year} {2009})},\ \Eprint
  {https://arxiv.org/abs/0908.4470} {arXiv:0908.4470 [hep-ph]} \BibitemShut
  {NoStop}%
\bibitem [{\citenamefont {Eto}\ \emph {et~al.}(2010)\citenamefont {Eto},
  \citenamefont {Nitta},\ and\ \citenamefont
  {Yamamoto}}]{EtoNittaYamamoto2010}%
  \BibitemOpen
  \bibfield  {author} {\bibinfo {author} {\bibfnamefont {M.}~\bibnamefont
  {Eto}}, \bibinfo {author} {\bibfnamefont {M.}~\bibnamefont {Nitta}},\ and\
  \bibinfo {author} {\bibfnamefont {N.}~\bibnamefont {Yamamoto}},\ }\href
  {https://doi.org/10.1103/PhysRevLett.104.161601} {\bibfield  {journal}
  {\bibinfo  {journal} {Phys. Rev. Lett.}\ }\textbf {\bibinfo {volume} {104}},\
  \bibinfo {pages} {161601} (\bibinfo {year} {2010})},\ \Eprint
  {https://arxiv.org/abs/0912.1352} {arXiv:0912.1352 [hep-ph]} \BibitemShut
  {NoStop}%
\bibitem [{\citenamefont {Gorsky}\ \emph {et~al.}(2011)\citenamefont {Gorsky},
  \citenamefont {Shifman},\ and\ \citenamefont {Yung}}]{GorskyShifmanYung2011}%
  \BibitemOpen
  \bibfield  {author} {\bibinfo {author} {\bibfnamefont {A.}~\bibnamefont
  {Gorsky}}, \bibinfo {author} {\bibfnamefont {M.}~\bibnamefont {Shifman}},\
  and\ \bibinfo {author} {\bibfnamefont {A.}~\bibnamefont {Yung}},\ }\href
  {https://doi.org/10.1103/PhysRevD.83.085027} {\bibfield  {journal} {\bibinfo
  {journal} {Phys. Rev. D}\ }\textbf {\bibinfo {volume} {83}},\ \bibinfo
  {pages} {085027} (\bibinfo {year} {2011})},\ \Eprint
  {https://arxiv.org/abs/1101.1120} {arXiv:1101.1120 [hep-ph]} \BibitemShut
  {NoStop}%
\bibitem [{\citenamefont {Eto}\ \emph {et~al.}(2011)\citenamefont {Eto},
  \citenamefont {Nitta},\ and\ \citenamefont
  {Yamamoto}}]{EtoNittaYamamoto2011}%
  \BibitemOpen
  \bibfield  {author} {\bibinfo {author} {\bibfnamefont {M.}~\bibnamefont
  {Eto}}, \bibinfo {author} {\bibfnamefont {M.}~\bibnamefont {Nitta}},\ and\
  \bibinfo {author} {\bibfnamefont {N.}~\bibnamefont {Yamamoto}},\ }\href
  {https://doi.org/10.1103/PhysRevD.83.085005} {\bibfield  {journal} {\bibinfo
  {journal} {Phys. Rev. D}\ }\textbf {\bibinfo {volume} {83}},\ \bibinfo
  {pages} {085005} (\bibinfo {year} {2011})},\ \Eprint
  {https://arxiv.org/abs/1101.2574} {arXiv:1101.2574 [hep-ph]} \BibitemShut
  {NoStop}%
\bibitem [{\citenamefont {Nitta}\ \emph {et~al.}(2014)\citenamefont {Nitta},
  \citenamefont {Uchino},\ and\ \citenamefont {Vinci}}]{NittaUchinoVinci2014}%
  \BibitemOpen
  \bibfield  {author} {\bibinfo {author} {\bibfnamefont {M.}~\bibnamefont
  {Nitta}}, \bibinfo {author} {\bibfnamefont {S.}~\bibnamefont {Uchino}},\ and\
  \bibinfo {author} {\bibfnamefont {W.}~\bibnamefont {Vinci}},\ }\href
  {https://doi.org/10.1007/JHEP09(2014)098} {\bibfield  {journal} {\bibinfo
  {journal} {JHEP}\ }\textbf {\bibinfo {volume} {09}},\ \bibinfo {pages}
  {098}},\ \Eprint {https://arxiv.org/abs/1311.5408} {arXiv:1311.5408 [hep-th]}
  \BibitemShut {NoStop}%
\bibitem [{\citenamefont {Tong}(2006)}]{TongSC}%
  \BibitemOpen
  \bibfield  {author} {\bibinfo {author} {\bibfnamefont {D.}~\bibnamefont
  {Tong}},\ }\href {https://doi.org/10.1088/1126-6708/2006/12/051} {\bibfield
  {journal} {\bibinfo  {journal} {JHEP}\ }\textbf {\bibinfo {volume} {12}},\
  \bibinfo {pages} {051}},\ \Eprint {https://arxiv.org/abs/hep-th/0610214}
  {arXiv:hep-th/0610214} \BibitemShut {NoStop}%
\bibitem [{\citenamefont {Affleck}\ and\ \citenamefont
  {Haldane}(1987)}]{AffleckHaldane1987}%
  \BibitemOpen
  \bibfield  {author} {\bibinfo {author} {\bibfnamefont {I.}~\bibnamefont
  {Affleck}}\ and\ \bibinfo {author} {\bibfnamefont {F.~D.~M.}\ \bibnamefont
  {Haldane}},\ }\href {https://doi.org/10.1103/PhysRevB.36.5291} {\bibfield
  {journal} {\bibinfo  {journal} {Phys. Rev. B}\ }\textbf {\bibinfo {volume}
  {36}},\ \bibinfo {pages} {5291} (\bibinfo {year} {1987})}\BibitemShut
  {NoStop}%
\bibitem [{\citenamefont {Itoi}\ and\ \citenamefont
  {Mukaida}(1994)}]{ItoiMukaida1994}%
  \BibitemOpen
  \bibfield  {author} {\bibinfo {author} {\bibfnamefont {C.}~\bibnamefont
  {Itoi}}\ and\ \bibinfo {author} {\bibfnamefont {H.}~\bibnamefont {Mukaida}},\
  }\href {https://doi.org/10.1088/0305-4470/27/14/011} {\bibfield  {journal}
  {\bibinfo  {journal} {J. Phys. A}\ }\textbf {\bibinfo {volume} {27}},\
  \bibinfo {pages} {4695} (\bibinfo {year} {1994})}\BibitemShut {NoStop}%
\bibitem [{\citenamefont {Affleck}\ \emph {et~al.}(1989)\citenamefont
  {Affleck}, \citenamefont {Gepner}, \citenamefont {Schulz},\ and\
  \citenamefont {Ziman}}]{AffleckGepnerSchulzZiman1989}%
  \BibitemOpen
  \bibfield  {author} {\bibinfo {author} {\bibfnamefont {I.}~\bibnamefont
  {Affleck}}, \bibinfo {author} {\bibfnamefont {D.}~\bibnamefont {Gepner}},
  \bibinfo {author} {\bibfnamefont {H.~J.}\ \bibnamefont {Schulz}},\ and\
  \bibinfo {author} {\bibfnamefont {T.}~\bibnamefont {Ziman}},\ }\href
  {https://doi.org/10.1088/0305-4470/22/5/015} {\bibfield  {journal} {\bibinfo
  {journal} {J. Phys. A}\ }\textbf {\bibinfo {volume} {22}},\ \bibinfo {pages}
  {511} (\bibinfo {year} {1989})}\BibitemShut {NoStop}%
\bibitem [{\citenamefont {Lieb}\ \emph {et~al.}(1961)\citenamefont {Lieb},
  \citenamefont {Schultz},\ and\ \citenamefont {Mattis}}]{LSM1961}%
  \BibitemOpen
  \bibfield  {author} {\bibinfo {author} {\bibfnamefont {E.~H.}\ \bibnamefont
  {Lieb}}, \bibinfo {author} {\bibfnamefont {T.}~\bibnamefont {Schultz}},\ and\
  \bibinfo {author} {\bibfnamefont {D.}~\bibnamefont {Mattis}},\ }\href
  {https://doi.org/10.1016/0003-4916(61)90115-4} {\bibfield  {journal}
  {\bibinfo  {journal} {Annals Phys.}\ }\textbf {\bibinfo {volume} {16}},\
  \bibinfo {pages} {407} (\bibinfo {year} {1961})}\BibitemShut {NoStop}%
\bibitem [{\citenamefont {Oshikawa}(2000)}]{Oshikawa2000}%
  \BibitemOpen
  \bibfield  {author} {\bibinfo {author} {\bibfnamefont {M.}~\bibnamefont
  {Oshikawa}},\ }\href {https://doi.org/10.1103/PhysRevLett.84.1535} {\bibfield
   {journal} {\bibinfo  {journal} {Phys. Rev. Lett.}\ }\textbf {\bibinfo
  {volume} {84}},\ \bibinfo {pages} {1535} (\bibinfo {year}
  {2000})}\BibitemShut {NoStop}%
\bibitem [{\citenamefont {Tasaki}(2022)}]{Tasaki2022}%
  \BibitemOpen
  \bibfield  {author} {\bibinfo {author} {\bibfnamefont {H.}~\bibnamefont
  {Tasaki}},\ }\href@noop {} {\  (\bibinfo {year} {2022})},\ \Eprint
  {https://arxiv.org/abs/2202.06243} {arXiv:2202.06243 [cond-mat.stat-mech]}
  \BibitemShut {NoStop}%
\bibitem [{\citenamefont {Hayata}\ and\ \citenamefont
  {Yamamoto}(2015)}]{HayataYamamoto2015}%
  \BibitemOpen
  \bibfield  {author} {\bibinfo {author} {\bibfnamefont {T.}~\bibnamefont
  {Hayata}}\ and\ \bibinfo {author} {\bibfnamefont {A.}~\bibnamefont
  {Yamamoto}},\ }\href {https://doi.org/10.1103/PhysRevA.92.043628} {\bibfield
  {journal} {\bibinfo  {journal} {Phys. Rev. A}\ }\textbf {\bibinfo {volume}
  {92}},\ \bibinfo {pages} {043628} (\bibinfo {year} {2015})},\ \Eprint
  {https://arxiv.org/abs/1411.5195} {arXiv:1411.5195 [cond-mat.quant-gas]}
  \BibitemShut {NoStop}%
\bibitem [{\citenamefont {Yamamoto}(2018)}]{Yamamoto2018}%
  \BibitemOpen
  \bibfield  {author} {\bibinfo {author} {\bibfnamefont {A.}~\bibnamefont
  {Yamamoto}},\ }\href {https://doi.org/10.1093/ptep/pty106} {\bibfield
  {journal} {\bibinfo  {journal} {PTEP}\ }\textbf {\bibinfo {volume} {2018}},\
  \bibinfo {pages} {103B03} (\bibinfo {year} {2018})},\ \Eprint
  {https://arxiv.org/abs/1804.08051} {arXiv:1804.08051 [hep-lat]} \BibitemShut
  {NoStop}%
\bibitem [{\citenamefont {Yamamoto}\ and\ \citenamefont
  {Hirono}(2013)}]{YamamotoHirono2013}%
  \BibitemOpen
  \bibfield  {author} {\bibinfo {author} {\bibfnamefont {A.}~\bibnamefont
  {Yamamoto}}\ and\ \bibinfo {author} {\bibfnamefont {Y.}~\bibnamefont
  {Hirono}},\ }\href {https://doi.org/10.1103/PhysRevLett.111.081601}
  {\bibfield  {journal} {\bibinfo  {journal} {Phys. Rev. Lett.}\ }\textbf
  {\bibinfo {volume} {111}},\ \bibinfo {pages} {081601} (\bibinfo {year}
  {2013})},\ \Eprint {https://arxiv.org/abs/1303.6292} {arXiv:1303.6292
  [hep-lat]} \BibitemShut {NoStop}%
\bibitem [{\citenamefont {Blote}\ \emph {et~al.}(1986)\citenamefont {Blote},
  \citenamefont {Cardy},\ and\ \citenamefont
  {Nightingale}}]{BloeteCardyNightingale1986}%
  \BibitemOpen
  \bibfield  {author} {\bibinfo {author} {\bibfnamefont {H.~W.~J.}\
  \bibnamefont {Blote}}, \bibinfo {author} {\bibfnamefont {J.~L.}\ \bibnamefont
  {Cardy}},\ and\ \bibinfo {author} {\bibfnamefont {M.~P.}\ \bibnamefont
  {Nightingale}},\ }\href {https://doi.org/10.1103/PhysRevLett.56.742}
  {\bibfield  {journal} {\bibinfo  {journal} {Phys. Rev. Lett.}\ }\textbf
  {\bibinfo {volume} {56}},\ \bibinfo {pages} {742} (\bibinfo {year}
  {1986})}\BibitemShut {NoStop}%
\end{thebibliography}%

\section*{End Matter}
\appendix
\section*{Exact metric identity}

For completeness, set $x=\mu_B r$.  Equation~(\ref{eq:alpha}) becomes
\begin{equation}
 \alpha_{xx}+{\alpha_x\over x}
 -{\nu^2\over x^2}\sin\alpha\cos\alpha
 +\sin\alpha\cos\alpha=0 .
 \label{eq:alphaX}
\end{equation}
Define
\begin{equation}
 {\cal P}(x)={x^2\over2}\alpha_x^2
 -{\nu^2\over2}\sin^2\alpha-{x^2\over2}\cos^2\alpha .
\end{equation}
Using Eq.~(\ref{eq:alphaX}) one finds
\begin{equation}
 {d{\cal P}\over dx}=-x\cos^2\alpha .
\end{equation}
Regularity gives ${\cal P}(0)=0$, while
$\alpha(\infty)=\pi/2$ and exponential approach to the vacuum give
${\cal P}(\infty)=-\nu^2/2$.  Hence
\begin{equation}
 \int_0^\infty x\,dx\,\cos^2\alpha
 ={\cal P}(0)-{\cal P}(\infty)={\nu^2\over2},
\end{equation}
which proves Eq.~(\ref{eq:exactmetric}).  For $\nu=1$ our numerical solution gives
$0.4999999996$.

\end{document}